\documentclass[a4paper,12pt]{article}
\pdfoutput=1
\usepackage{amsmath}
\usepackage{graphicx}
\usepackage{amsfonts}
\usepackage{amssymb}
\usepackage{color}

\usepackage[bookmarks=true,colorlinks,unicode]{hyperref} 
\def\equationautorefname~#1\null{Eq.(#1)\null}
\def\figureautorefname~#1\null{Fig.#1\null}

\newcommand{\bc}{\begin{center}}
\newcommand{\ec}{\end{center}}
\newcommand{\beq}{\begin{equation}}
\newcommand{\eeq}{\end{equation}}
\newcommand{\beqa}{\begin{eqnarray}}
\newcommand{\eeqa}{\end{eqnarray}}
\newcommand{\beqs}{\begin{eqnarray*}}
\newcommand{\eeqs}{\end{eqnarray*}}

\newcommand{\bi}{\begin{itemize}}
\newcommand{\ei}{\end{itemize}}

\newcommand{\half}{\frac{1}{2}}

\newcommand{\denisZ}{\mathcal{Z}}

\newtheorem{lem}{Lemma}

\newtheorem{cor}{Corollary} 

\newtheorem{prop}{Proposition}

\newtheorem{conj}{Conjecture}

\newtheorem{defi}{Definition}

\title{\bf Stochastic spikes and strong noise limits \\ of stochastic differential equations}
\author{}
\date{}

\begin{document}
\maketitle
\pagestyle{empty}

\vskip -0.5 truecm

\centerline{M. Bauer${}^{\spadesuit,\diamondsuit}$\footnote{michel.bauer@cea.fr} and D. Bernard${}^{\clubsuit}$\footnote{denis.bernard@ens.fr}}

\vskip 0.5 truecm

\noindent
\small{${}^\spadesuit$ Institut de Physique Th\'eorique de Saclay, CEA-Saclay $\&$ CNRS, 91191 Gif-sur-Yvette, France.}\\
\small{${}^\diamondsuit$ D\'epartement de math\'ematiques et applications, \'Ecole Normale Sup\'erieure, PSL Research University, 75005 Paris, France}\\
\small{$^\clubsuit$ Laboratoire de Physique Th\'eorique de l'\'Ecole Normale Sup\'erieure de Paris, CNRS, ENS $\&$ PSL Research University, UPMC $\&$ Sorbonne Universit\'es, France.}
\vskip 0.5 truecm

\small{\centerline{\today}}
\vskip 1.0 truecm

\pagestyle{plain}

\begin{abstract}
Motivated by studies of indirect measurements in quantum mechanics, we investigate stochastic differential equations with a fixed point subject to an additional infinitesimal repulsive perturbation. We conjecture, and prove for an important class, that the solutions exhibit a universal behavior when time is rescaled appropriately: by fine-tuning of the time scale with the infinitesimal repulsive perturbation, the trajectories converge in a precise sense to spiky trajectories that can be reconstructed from an auxiliary time-homogeneous Poisson process. 

Our results are based on two main tools. The first is a time change followed by an application of Skorokhod's lemma. We prove an effective approximate version of this lemma  of independent interest. The second is an analysis of first passage times, which shows a deep interplay between scale functions and invariant measures. 

We conclude with some speculations of possible applications of the same techniques in other areas.

\end{abstract}

\newpage

\section{Introduction}
\label{sec:intro}

We are interested in the strong noise limit for a class of stochastic differential equations (SDEs) of a specific form whose solutions develop spiky behaviors which may be precisely described.
We are looking at SDEs for a single positive variable $X_t$, driven by a Brownian motion $B_t$, of the following form :
\beq \label{sde:general}
 dX_t = \frac{\lambda^2}{2}\big(\epsilon  a(X_t) -  b(X_t)\big)\, dt + \lambda c(X_t)\, dB_t ,
 \eeq
when the coefficients $a,b,c$ satisfy certain conditions and $\epsilon,\lambda$ are free parameters. We assume that $x=0$ is the only fixed point of the equation at $\epsilon=0$ and that it is attractive. This implies in particular that $b(0)=c(0)=0$, and adding that the function $b$ is non-negative is a sufficient --but not necessary, see below-- condition. We also impose that whenever $\epsilon >0$ (even very small) the contribution of $a$ is strong enough to prevent trajectories from reaching $0$, but that the neighborhood of $0$ is the only region where the $\epsilon a$ term plays a significant role. We shall make these constraints quantitative later, at least they impose that $a(x) \geq 0$ close to  $x=0$.

We are interested in the limit $\lambda\to \infty$, with an appropriate scaling of $\epsilon>0$ as $\lambda\to\infty$ for the limit to be meaningful. With the choices we made for the parameter dependence, $\lambda$ can be reabsorbed by a simple rescaling of time  time and we can interpret the limit we look for as the $\epsilon \to 0^+$ limit but with a tuning of the time scale (depending on $\epsilon$) that ensures that certain features (some first passage times in the case at hand) remain finite. 

However, as we shall briefly exemplify below, SDEs of this form arise in the description of the evolution of quantum systems under continuous monitoring and then the natural interpretation is different, dual in some sense. The parameter $\lambda$ is directly related to the rate at which information is pulled out from the systems (recall that extracting information from a quantum system induces a random back-action), and the other natural physical parameter is $\eta=\lambda^2\epsilon/2$ so that \autoref{sde:general} could be rewritten as
\beq 
dX_t = \big(\eta a(X_t) - \frac{\lambda^2}{2} b(X_t)\big)\, dt + \lambda\, c(X_t)\, dB_t ,
\eeq
where the terms involving $\lambda$ are due to measurements and bring in randomness, while the term involving $\eta$ corresponds to a standard, deterministic, evolution. Larger and larger $\lambda$s means more and more measurements per unit time, and the 
the way $\epsilon$ --or equivalently $\eta$-- has to be scaled with $\lambda$ is just a quantitative expression of the quantum Zeno effect for the problem at hand. As physicists, this is our main motivation and this is why we interpret our study of \autoref{sde:general} as a strong noise limit.

The prototype example is provided by the following SDE, with $b > -1$,
\beq \label{sde:b0}
dX_t = \frac{\lambda^2}{2}\,(\epsilon - b X_t)dt + \lambda\, X_t\, dB_t,
\eeq
in the limit $\epsilon\to 0^+$, $\lambda\to\infty$ with the product $\lambda^2\epsilon^{b+1}:=J$ fixed. The parameter $\epsilon$ is dimensionless and $\lambda^{2}$ has the dimension of a frequency (inverse of time). 
This SDE codes for two effects: (i) the net effect of $\epsilon$-independent terms is to attract $X_t$ to the origin (even if $b \in ]-1,0[$!) on a time scale of order $\lambda^{-2}$ and (ii) the terms proportional to $\epsilon$ pushes $X_t$ away from the origin. The latter effect is comparatively active only when $X_t$ is close enough to the origin. These effects are competing in the combined limit $\epsilon\to 0$, $\lambda\to\infty$ and they give rise to a non-trivial behavior provided $\epsilon$ and $\lambda$ are scaled in an appropriate way. It is easy to verify that, at large $\lambda$ and fixed $t$, $X_t$ is of order $\epsilon$. More precisely  there is an explicit distribution function $F$ such that for, each fixed $t$, $\lim_{\lambda \to \infty} \text{Prob}(X_t < \epsilon y)=F(y)$. 
However, the process $t\to X_t$ is non-trivial in the scaling limit $\lambda\to\infty$ with $\lambda^2\epsilon^{b+1}=J$ fixed. Indeed, what is true at any time does not hold at every time. To see this, let us ask ourselves what is the distribution of the $X$-maxima in the time interval $[0,T]$. Since at $\epsilon \simeq 0$ the typical time scale of this process is $\lambda^{-2}$, we make the approximation that the portions of the $X$-trajectories separated by a time laps bigger than $\lambda^{-2}$ are independent. The probability that the $X$-maxima is less than a given number $m$ in the time interval $[0,T]$ may then be estimated as
\[ \mathbb{P}\big[X_t <m,\ t\in[0,T]\big] \simeq \Big[ \int_0^m P_\mathrm{inv}(x)\, dx \Big]^{\lambda^2 T} \]
with $P_\mathrm{inv}(x)\, dx=  \frac{\epsilon^{b+1}}{\Gamma(b+1)}\, \frac{dx}{x^{b+2}}\, e^{-\epsilon/x }$ the invariant measure of \autoref{sde:b0}. The important point is the heavy tail of this measure. Hence,
 \[ 
\mathbb{P}\big[X_t <m,\ t\in[0,T]\big] \simeq  \Big[ 1 - \mathrm{const.}\, \frac{\epsilon^{b+1} }{ m^{b+1} }\Big]^{\lambda^2 T} \simeq e^{-\mathrm{const.} (\lambda^2 \epsilon^{b+1})\, T/m^{b+1} }.
\]
We thus get a non trivial distribution for the $X$-maxima in the scaling limit $\lambda^2 \epsilon^{b+1}=J$ fixed, although, for each given $t$, $X_t$ is almost surely zero. These seemingly self-contradictory statements actually mean that, on any time interval $[0,T]$, the process $t\to X_t$ makes series of excursions -- instantaneous in the limit $\lambda\to\infty$ and which we call `spikes' -- away from the origin. These excursions form the stochastic spikes. 
See \autoref{fig:linear_spikes} in \autoref{ssec:slpt} for samples exhibiting the dependence in the parameter $b$.

One of the aims of this paper is to present a detailed characterization of the large noise scaling limit of solutions of \autoref{sde:b0} and their spiky behavior:

\begin{prop}
\label{prop:exactsolv} 
	In the scaling limit $\lambda\to\infty$, $\epsilon\to0$ with $\lambda^2\epsilon^{b+1}=:J$ fixed, the solution $X_t$ of the SDE (\ref{sde:b0}), $dX_t = \frac{\lambda^2}{2}\,(\epsilon - b X_t)dt + \lambda\, X_t\, dB_t$, is a fractional power of a reflected Brownian motion parametrized by its local time. Namely, in law and in the scaling limit,
\beq \label{Xt-local}
 X_t = (b+1)^\frac{1}{b+1}\ |\tilde W_\tau|^\frac{1}{b+1},\quad J\, t = 2\Gamma(b+1)\ L_\tau,
\eeq
with $L_\tau$ the local time at the origin of a standard Brownian motion $\tilde W_\tau$.
\end{prop} 

Because the local time stays constant when $\tilde W_\tau$ is away from zero, any excursion of $\tilde W_\tau$ away from the origin is mapped to an instantaneous spiky excursion of $X_t$. This has a direct consequence for the distribution of the heights of the $X$-spikes:
 
\begin{cor}	
\label{cor:spikexactsolv} 	
	The tips of the $X$-spikes form a  Poisson point process on $\mathbb{R}\times \mathbb{R}_+$ with intensity
\[  d\nu = \frac{(b+1)}{2\Gamma(b+1)}\, Jdt\, \frac{dx}{x^{b+2}}.\]
\end{cor}

The proof of this proposition is presented in \autoref{sec:proofs}. It is based on using a Skorokhod decomposition of a process $Q_t$, defined in terms of $X_t$, to relate $X_t$ to the power of a Brownian motion -- see  \cite{skorokhod,yenyor} for an introduction to Skorokhod's decompositions --  and on using an ergodicity theorem to relate the original time $t$ to the effective local time $L_\tau$. We shall also give an independent check that the $X$-spikes form a Poisson point process by determining, in the scaling limit, the distribution of a family of stopping times $T_{y\to z}$ defined as the times for $X_t$ to go from $y$ to $z$.

In  \autoref{sec:more} we shall extend these results to a larger class of SDEs of the form
\beq\label{sde:more}
 dX_t = \frac{\lambda^2}{2}\,(\epsilon X_t^q - b X_t^n)dt + \lambda\, X_t^k\, dB_t,
 \eeq
with $n=2k-1>q\geq 0$ and $b >-1$. By a simple change of variable, namely $Q_t:= \frac{X_t^{b+1}}{b+1}$, this equation can be rewritten as
\beq \label{sde:autre}
 dQ_t = \frac{\hat \lambda^2}{2} \hat \epsilon\, Q_t^\alpha dt + \hat \lambda\, Q_t^\delta dB_t,
 \eeq
with $\alpha=\frac{b+q}{b+1} $ and $\delta=\frac{b+k}{b+1}$ and with $\hat \lambda$ and $\hat \epsilon$ respectively proportional to $\lambda$ and $\epsilon$. The condition $2k-1>q>0$ translates into $2\delta-1>\alpha>0$. In other words, it is equivalent to study the SDEs of the form (\ref{sde:more}) or (\ref{sde:autre}) but the former arise more naturally in some physical problems (see below).  The large noise limit we shall consider is $\lambda\to\infty$, $\epsilon\to0$ with $\lambda^2\, \epsilon^\frac{b+n}{n-q}=:J$ fixed, or equivalently $\hat \lambda^2 \hat \epsilon^\frac{2\delta-1}{2\delta-1-\alpha}=: \hat J$ fixed:

\begin{prop}
    \label{prop:moregeneral}
  In the scaling limit $\lambda\to\infty$, $\epsilon\to0$ with $\lambda^2\epsilon^\frac{b+n}{n-q}=:J$ fixed, the solution $X_t$ of the SDE (\ref{sde:more}), with $n=2k-1>q$, is identical in law with a fractional power of a reflected Brownian motion parametrized by its local time:
\beq \label{Xt-local-bis}
 X_t = (b+1)^\frac{1}{b+1}\ |\tilde W_\tau|^\frac{1}{b+1},\quad J\, t = 2\denisZ\ L_\tau,
\eeq
with $L_\tau$ the local time at the origin of a standard Brownian motion $\tilde W_\tau$ and $\denisZ$ a numerical factor depending on the parameters $q,n,k$.
\end{prop}

The proof of \autoref{prop:moregeneral}  is parallel to that of \autoref{prop:exactsolv} and we shall give only the main ingredients omitting the details. We choose to split the two propositions and to deal first with the particular cases of \autoref{prop:exactsolv}  (which correspond to $q=0$, $n=1=k$) in order to make the proof more readable.

We shall also outline in \autoref{sec:general} the expected results for the general class of SDEs of the form (\ref{sde:general}), including (\ref{sde:more}) with $n\not=2k-1$. These results are presented as motivated conjectures. There are yet a few delicate points to extend the previous proofs to the general case. In particular, the proof of the reconstruction relation between the effective local time $L_\tau$ and the original time uses the ergodic theorem which can only be rigorously applied only if the SDE possesses some specific properties which allow to disentangle the scaling limit $\lambda\to\infty$, $\epsilon\to 0$. 

A brief comparison with weak noise limit is sketched in \autoref{sec:weak}

\section{Motivation from Quantum Mechanics}
\label{sec:MQ}

Our motivation for characterizing the strong noise limit of solutions of SDEs of the form (\ref{sde:general}) find its roots in the study of quantum systems under continuous monitoring. Let $\rho_t$ be the state (density matrix) of the quantum systems. Under continuous monitoring, its time evolution is governed by a SDE of the following form:
\[ d\rho_t = L_\mathrm{sys}(\rho_t)\, dt + L_N(\rho_t)\, dt + M_N(\rho_t)\, dB_t,\]
where $L_\mathrm{sys}$ is a Lindbladian describing the system evolution in absence of monitoring, while $L_N(\rho):= N^\dag\rho N -\half(NN^\dag\rho+\rho NN^\dag)$ and $M_N(\rho):= N^\dag\rho+\rho N - \rho\, \mathrm{Tr}(N^\dag\rho+\rho N)$ for some operator $N$ specifying the way the monitoring is implemented. Solutions of these SDEs are called quantum trajectories.

We present here a few examples which lead to SDEs of the form (\ref{sde:general}). We shall deal with qubits whose Hilbert space is $\mathbb{C}^2$.
We parametrize the qubit density matrix by $\rho=\half(\mathbb{I}+ \vec{\sigma}\cdot\vec{S})$ where $\sigma^{x,y,z}$ are the standard Pauli matrices and $\vec{S}^2\leq 1$.  Setting $2P:=1+S^z$ and $2U:=S^x-iS^y$. We have:
\[ \rho=\half \begin{pmatrix} 1+S^z & S^x-iS^y\\ S^x+iS^y & 1-S^z \end{pmatrix} = \begin{pmatrix} P & U\\ U^* & 1-P \end{pmatrix}.\]

\subsection{QND detection of a thermal qubit}

This case was studied in detail in \cite{BBTskospikes}.

\begin{figure}
	\begin{center} \includegraphics[width=0.90\textwidth]{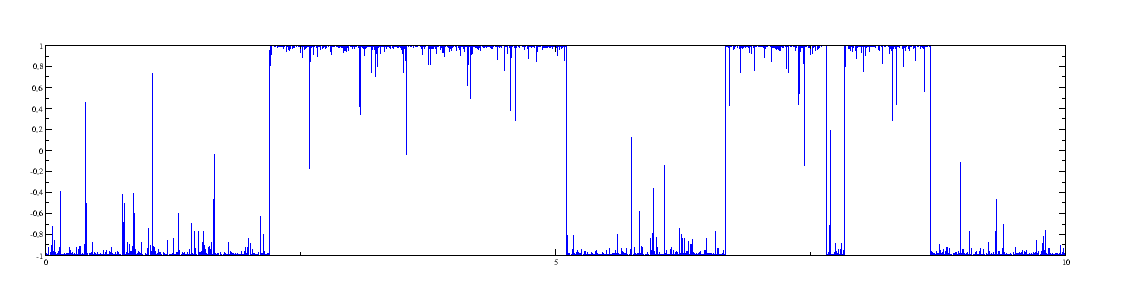} \end{center}
	\caption{\emph{A typical sample of thermal jumps and spikes. The ``curve'' represents the probability to be in the $+$ $\sigma^z$ eigenstate}.}
	\label{fig:thermal_jumps_and_spikes}
\end{figure}

Let us suppose that the qubit energy eigenstates are the $\sigma^z$ eigenstates. When the energy is continuously measured, $U$ decouples and the evolution is governed by the equation: 
\[ dP_t= \eta (p- P_t) dt + \lambda P_t(1-P_t)\, dB_t .\]
Efficient monitoring corresponds to a large value of $\lambda$, and $p$ is the probability to be in the $+$  $\sigma^z$ eigenstate at thermal equilibrium. 
A typical sample at fixed $\eta$ and $p$ but large $\lambda$ is presented in \autoref{fig:thermal_jumps_and_spikes}. 

The linearized version for $P_t$ small reads (after renaming $P_t$ to $X_t$) : 
\[ dX_t = (\eta p)\, dt + \lambda X_t\, dB_t \]
The scaling  is $\lambda\to\infty$ with $\eta p$ fixed. 
A typical sample  is presented in \autoref{ssec:slpt}, see \autoref{fig:linear_spikes}. 

\subsection{QND detection of Rabi oscillation} 

Rabi oscillation refers to the precession of a spin one-half in an external magnetic field. Let suppose that the magnetic field is along the $y$-axis and that the spin along the $z$-axis is monitored using non-demolition measurements (this corresponds to $N\propto \sigma^z$). Then the quantum trajectory SDEs read:
\begin{eqnarray*}
 dP_t &=& \omega\, U_t\, dt + \gamma P_t(1-P_t)\, dB_t,\\
 dU_t &=& -\omega\, (P_t-\frac{1}{2})\,dt - \frac{ \gamma^2}{8} U_t\, dt - \gamma U_t(P_t -\frac{1}{2})\,dB_t
 \end{eqnarray*}
 with $\omega$ the so-called Rabi frequency proportional to the magnetic field.  Here $\gamma^{-2}$ codes for the information extraction rate. These equations preserve pure states (actually any initial state converges exponentially to a pure state). We thus restrict ourselves to pure states of the form
$ |\psi_t \rangle= \cos(\theta_t/2) |+\rangle- \sin(\theta_t/2) |-\rangle$, with $|\pm\rangle$ the state basis of spin pointing up or down in the $z$-direction i.e. $\sigma^z|\pm\rangle=\pm |\pm\rangle$. The SDE then reduces to: 
\[  d\theta_t= (\omega - 2\gamma^2 \sin\theta_t\cos\theta_t)\, dt - 2\gamma \sin \theta_t\, dB_t.\]
The linearized version is (after renaming $\theta_t$ to $X_t$ and $B_t$ to $-B_t$)
\[ dX_t= \frac{\lambda^2}{2}(\epsilon-X_t)\, dt + \lambda X_t\, dB_t,\]
where $\lambda:=2\gamma$ and $\omega:=2\gamma^2\epsilon$.

In order for the (nonlinear or linearized) equations to have a non-trivial limit when the information extraction rate is large, one must impose a scaling relation $\lambda^2\epsilon^2=:J$ fixed, which translates into $\omega\sim \lambda$. The need for rescaling the Rabi frequency is an echo of the Zeno effect. Indeed, in the large $\gamma$ limit, the monitoring back-action induces quantum jumps separated by time intervals of order $\bar T_\mathrm{jump}=\gamma^2/\omega^2$. The scaling relation simply imposes the finiteness of this mean time interval since $\bar T_\mathrm{jump}\propto J^{-1}$.

\begin{figure}
	\begin{center}\includegraphics[width=0.95\textwidth]{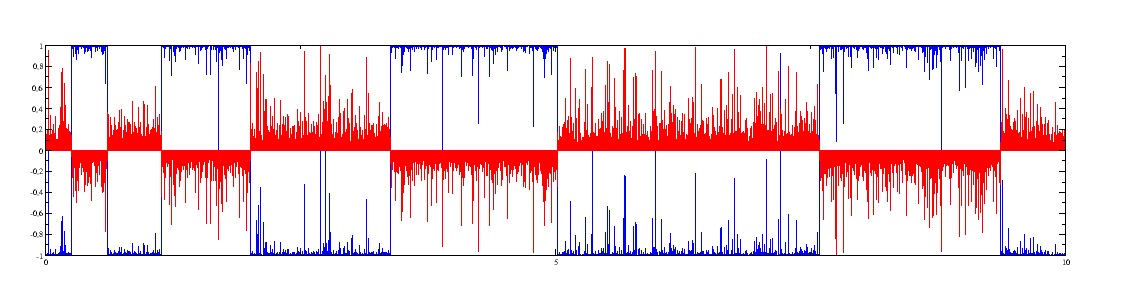}\end{center}
	\caption{\emph{A typical sample of jumps and spikes induced by indirect measurements. The blue ``curve'' represents $S^z$, the red ``curve'' is $S^x$.}}
	\label{fig:rabi_jumps_and_spikes}
\end{figure}

A typical sample at large $\lambda$ with fixed $\lambda^2\epsilon^2$ is presented in \autoref{fig:rabi_jumps_and_spikes}. 

A typical sample  of the linearized version is presented in \autoref{ssec:slpt}, see \autoref{fig:linear_spikes}.

\subsection{Homodyne detection of Rabi oscillation}

Homodyne detection corresponds to monitor the qubit via a coupling $N\propto \sigma_+$ (recall $\sigma_\pm=(\sigma^x\pm i\sigma^y)/2$).

This procedure is nowadays routinely implemented in circuit QED. The output signal is drifted by the mean value of $\sigma^x=\sigma_++\sigma_-$ (We could also have considered a coupling via $i\sigma_+$ so that the signal would have been drifted by $\sigma^y$). In presence of Rabi oscillations at a frequency $\omega$ and for a magnetic field pointing in the $y$-direction, the homodyne quantum trajectory SDEs are
\beqs
dS^z_t &=& \omega S^x_t dt+ \gamma^2 (1-S^z_t) dt + \gamma (1-S^z_t)S^x_t dB_t \\
dS^x_t &=& -\omega S^z_t dt -\frac{\gamma^2}{2}S^x_t dt + \gamma( 1-S^z_t - {S^x_t}^2)dB_t\\
dS^y_t &=& -\frac{\gamma^2}{2}S^y_t dt + \gamma S^y_t S^x_t dB_t
\eeqs
Again $\gamma^{-2}$ codes for the information extraction rate. 
Although the homodyne output signal is coupled to $\sigma^x$, so that the measurement gives information on the mean value of $\sigma^x$, the homodyne process is quite different from the previous (non-demolition) monitoring of $\sigma^x$. 
Purity is preserved by the homodyne evolution. We set $S^z=\cos\theta$ and $S^x=-\sin\theta$ and $S^y=0$. Then
\[  d\theta_t = \big[\omega -\frac{\gamma^2}{2}\sin\theta_t(1+2\sin^2(\theta_t/2))\big]\,dt + 2\gamma\sin^2(\theta_t/2)\, dB_t \]
The ``linearized'' version\footnote{An abuse of language: by this we mean that only the most relevant terms are kept.} for small $\theta$ yields (after renaming  $X_t=\theta_t/2$, $\lambda=\gamma$ and $\omega=\lambda^2\epsilon$):
\[ dX_t = \frac{\lambda^2}{2}\,(\epsilon -X_t)\, dt + \lambda\, X_t^2\, dB_t .\] 
The difference with the previous example is the presence of an $X^2_t$ (not $X_t$) factor in the noise term. But this equation fits in the form (\ref{sde:general}) of the class of SDEs we study.

\section{An effective form of Skorokhod's lemma}

Our convergence results rely on a simple extension of Skorokhod's lemma with explicit bounds. 

\begin{defi} \label{defi:decomp}
	Let $f_t, t\in [0,+\infty[$ be a real continuous function such that $f_0\geq 0$. \\
	-- An admissible pair for $f$ is a pair $(b,y)$ where $b_t, t\in [0,+\infty[$ and $y_t, t\in [0,+\infty[$ are real continuous functions, $b$ is nondecreasing, $y$ is nonnegative, $y=b+f$, and $y_0=f_0$. \\
	-- A Skorokhod decomposition for $f$ is an admissible pair $(a,x)$ such that $a$ is locally constant on $\{x> 0\}$.\\
	-- Let $\beta,\upsilon > 0$. An approximate $(\beta,\upsilon)$ decomposition of $f$ is an admissible pair $(b,y)$ such that for any $0 \leq s \leq u$, if $y_t \geq \upsilon$ for $t\in [s,u]$ then $b_u-b_s \leq \beta (u-s)$.
\end{defi}

Let us comment these definitions: \\
-- If $(a,x)$ is a Skorokhod decomposition of $f$, the set $\{x\neq 0\}$ which is open by continuity, is a countable union of disjoint open intervals of $[0,+\infty[$ and the condition on $a$ is that it is constant on each of these intervals. As $a$ is nondecreasing, it defines a nonnegative measure $da$ and using that $x$ is nonnegative the condition can be seen to be equivalent to $\int_B x_tda_t=0$ on each Borel set $B$ of $[0,+\infty[$. This may be rephrased informally as: $a$ may only increase when $x$ vanishes. \\
-- It is the immediate that a  Skorokhod decomposition of $f$ is an approximate $(\beta,\upsilon)$ decomposition for every $\beta, \upsilon>0$. Conversely, in an approximate decomposition with $\beta,\upsilon$ small, $b$ may only increase  significantly when $y$ is small. 

We then have:

\begin{lem} (The approximate Skorokhod decomposition)\label{lem:approsko} \\ 
-- Each real continuous function $f_t, t\in [0,+\infty[$ such that $f_0\geq 0$ has a  unique Skorokhod decomposition $(a,x)$, and $a$ is given explicitly as $a_t=\max(0,\max_{s \leq t} -f_s)$. \\
-- If $(b,y)$ is an admissible pair for $f$ then $b-a=y-x$ is a nonnegative function.\\
-- If $(b,y)$ is an approximate $(\beta,\upsilon)$ decomposition of $f$, then $0 \leq b_t-a_t \leq \upsilon + \beta t$.
\end{lem}

The first assertion is nothing but Skorokhod's original lemma which is elementary, see \cite{skorokhod,yenyor}. We shall reprove it together with the other assertions which are as elementary but seem to be new. 

{\bf Proof:} We first prove the existence part of the original Skorokhod lemma.\\ 
Let $m_t:= \max_{s \leq t} \{-f_s\}$ and note that $m+f$ is a nonnegative continuous function. We check that $a_t:=\max \{0,m_s\}$, $x_t:=a_t+f_t$ is a Skorokhod decomposition of $f$.  First the fact that $a$ is continuous and increasing is clear because so is $m$. Then $x$ is also continuous. We rewrite $x$ as $x_t=\max \{f_t,m_t+f_t\}\geq m_t+f_t$. As $m_t+f_t \geq 0$, $x$ is nonnegative. Suppose $t$ is such that $x_t>0$. Either $m_t <0$  and then by continuity $m_u <0$ for $u$ close to $t$ so that $a$ is $0$ (and in particular constant) close to $t$. Or $m_t \geq 0$, and then  $0 < x_t=m_t+f_t$, so $m_t > -f_t$, $m_u$ is constant and $\geq 0$ close to $t$ and $a$ is constant close to $t$. Thus $a$ is locally constant on $\{x> 0\}$. Finally, $a_0=\max \{0,-f_0\}=0$ because $f_0\geq 0$. So $x_0=f_0$. \\ 
Now we turn to the second assertion. We assume that $(a,x)$ is a Skorokhod decomposition of $f$ (not necessarily given by explicit formula above) and that $(b,y)$ is an admissible pair. We argue by contradiction\footnote{The same kind of argument is used several times below, and we call it a backtracking argument: to get a hold on something that happens at time $t$, we go backward in time up to a time when things were under control, and then propagate the control to the future up to $t$.}. Suppose there is some $t$ such that $a_t>b_t$. As $a_0=b_0=0$ we have  $t>0$ and $\bar{t}:=\sup\{0\leq s < t, b_s \geq a_s\}$ is well-defined. By continuity, $b_{\bar{t}}=a_{\bar{t}}$ and $0\leq y_s < x_s$ for $s\in ]\bar{t},t[$. Thus $a$ is constant on $]\bar{t},t[ $ and then, by continuity, on $[\bar{t},t]$. Hence \[0 < a_t-b_t=a_{\bar{t}}-b_t=(a_{\bar{t}}-b_{\bar{t}})+(b_{\bar{t}}-b_t)=b_{\bar{t}}-b_t.\]
This would contradict the fact that $b$ is nondecreasing. Thus $b-a=y-x$ is a nonnegative function, i.e. each piece of a Skorokhod decomposition is less than the corresponding piece in any admissible pair. As a Skorokhod decomposition is in particular an admissible pair, the Skorokhod decomposition is unique.\\
It remains to get the explicit bound for approximate $(\beta,\upsilon)$ decompositions. Thus let $(b,y)$ be an approximate $(\beta,\upsilon)$ decomposition of $f$, and $(a,x)$ be the Skorokhod decomposition. 
We first note that $0\leq b_t-a_t=y_t-x_t \leq  y_t$ (the first inequality was just proved above). 
Either $y_t \leq \upsilon$ and then $0\leq b_t-a_t\leq \upsilon$. Or $y_t > \upsilon$. This case splits again in two. Either $y_s \geq \upsilon$ for $s\in [0,t]$, and then $b_t-a_t\leq b_t=b_t-b_0\leq \beta t$. Or we can use backtracking and set $\bar{t}:= \sup\{0\leq s < t, y_s \leq \upsilon\}$, so that $y_{\bar{t}}=\upsilon$ and $y_s\geq \upsilon$ for $s\in [\bar{t},t]$. Then
\[b_t-a_t\leq b_t-a_s=(b_t-b_s)+(b_s-a_s) \leq (b_t-b_s)+ y_s =(b_t-b_s)+\upsilon \leq \beta (t-s)+\upsilon. \]
In all case, we have $0 \leq b_t -a_t \leq \beta t +\upsilon$, concluding the proof. $\square$

Note the immediate consequence of \autoref{lem:approsko}:

\begin{cor} \label{cor:skoconv}
	Suppose given a family of $(b^\varepsilon,y^\varepsilon)$ of $(\beta(\varepsilon),\upsilon(\varepsilon))$ decompositions of $f$, defined for $\epsilon >0$ and such that $\lim_{\varepsilon \to 0} \beta(\varepsilon)=\lim_{\varepsilon \to 0} \upsilon(\varepsilon)=0$. Then $(b^\varepsilon,y^\varepsilon)$ converges (uniformly on each compact interval) to the Skorokhod decomposition $(a,x)$ of when $\varepsilon \to 0$.
\end{cor}

We shall use these results for a special class of problems. \\
Let $\alpha >1$. Let $f$ be given and let $(a,x)$ be its Skorokhod decomposition. Assume that, for each $\varepsilon >0$, $y^\varepsilon$ is nonnegative and solves the integral equation 
\begin{equation} \label{eq:skoapprox} y^\varepsilon_t=f_t+\varepsilon \int_{0}^t \frac{du}{(y_u^\varepsilon)^\alpha}.\end{equation}
Setting $b^\varepsilon_t:=\varepsilon \int_{0}^t \frac{du}{(y_u^\varepsilon)^\alpha}$ it is immediate that $(b^\varepsilon,y^\varepsilon)$ is an approximate $(\varepsilon C^{-\alpha},C)$ decomposition of $f$ for any $C >0$. Optimizing on $C$, we have that $b^\varepsilon_t-a_t \leq (1+\frac{1}{\alpha})(\alpha \varepsilon t)^{\frac{1}{1+\alpha}}$, leading to the uniform convergence of $b^\varepsilon$ to $a$ on compact intervals. \\
In a typical application, $f$ will be a Brownian motion started somewhere on the positive axis. As $\alpha >1$, one sees (for instance by comparison with the $3d$ Bessel process) that a solution of \autoref{eq:skoapprox} exists, is unique and nonnegative, leading to the desired conclusions.

\section{Strong noise limit for a class of SDEs}
\label{sec:proofs}

We present here the different steps to prove \autoref{prop:exactsolv}.
Let us consider the SDE (\ref{sde:b0}) that we rewrite here :
\beq \label{sde:b0bis}
 dX_t = \frac{\lambda^2}{2}\,(\epsilon - b X_t)dt + \lambda\, X_t\, dB_t .
\eeq 
We assume $\epsilon>0$ and $b > -1$. 

We shall prove that the distribution of the solutions has a non trivial limit, called the scaling limit in what follows, when  $\lambda\to \infty$ and $\epsilon\to 0$ with $\lambda^2 \epsilon^{b+1}=:J$ fixed. We shall also describe the scaling limit in terms of Poisson point processes. 

Note that $\lambda^2\epsilon \ll \lambda^2$, so that the leading contribution away from the origin comes from the truncated SDE $dX_t = - \frac{\lambda^2}{2}\,b X_t\,dt + \lambda\, X_t\, dB_t$ where $\epsilon$ has been neglected. For $b \geq 0$ the solutions of the truncated equation are supermartingales bounded below by $0$ so they converge almost surely at large $t$. As $0$ is the sole fixed point, they do in fact converge almost surely to $0$ at large $t$. It turn out that these conclusions are valid more generally when $b > -1$ though in that case we rely on the explicit solution (at $\epsilon=0$) $X_t=X_0 e^{\lambda B_t-\frac{\lambda^2(b+1)}{2}t}$ for a proof. 

But whatever small $\epsilon$, when a solution $X_t$ to the complete equation comes very close to $0$ it is kicked away by the $\epsilon$-term and the scaling limit is chosen to ensure that these kicks are of the right order of magnitude. Much of the discussion that follows is an elaboration on this point. 

The complete equation \autoref{sde:b0bis} turns out to have a closed form solution (using a method based on "variation of constants"): 
\[ X_t= \Big[ X_0 + \frac{\lambda^2\epsilon}{2}\, \int_0^t du\, e^{-\lambda B_u+\frac{\lambda^2(b+1)}{2}u}\Big]\,e^{\lambda B_t-\frac{\lambda^2(b+1)}{2}t} .\]
For $\epsilon=0$, it is plain that the solutions converge to $0$ almost surely, in agreement with the general remark above. When $b>0$, the convergence also takes place in $L^1$. 
 
The invariant measure of \autoref{sde:b0bis} is the inverse $\Gamma$-distribution: $P_\mathrm{inv}(x)dx =  \frac{\epsilon^{b+1}}{\Gamma(b+1)}\, \frac{dx}{x^{b+2}}\, e^{-\epsilon/x }$, and $P_\mathrm{inv}(x)dx\to \delta_0$ in the limit $\epsilon\to 0$ as expected, confirming that $X_t$ converges to zero at large $t$ when $\epsilon=0$.  

We shall also make use of the following scaling property: writing $Y_t:=\epsilon^{-1} X_{t \lambda^{-2} }$ and noticing that $\tilde B_t:=\lambda B_{t\lambda^{-2}}$ is another Brownian motion, $Y_t$ solves an SDE independent of $\epsilon$ and $\lambda$, namely $dY_t= \frac{1}{2}(1-bY_t)dt+ Y_t d\tilde B_t$. In particular if one solves the SDE for an initial condition which is scale covariant (like $Y_0=0$, or $Y_0$ sampled with the stationary measure and independent of $\tilde B$) we can get a coupling of the corresponding solutions of the SDE (\ref{sde:b0bis}) for all values  of $\epsilon$ and $\lambda$.

\subsection{Scaling limit via time change}

It is clear (for instance from the explicit solution above) that if $X_0\geq 0$ then $X_t>0$ for $t >0$.  Let us change variable and set $Q_t:=\frac{X_t^{b+1}}{b+1}$. Then:
\[ dQ_t = \frac{\hat \lambda^2}{2}\, \hat \epsilon\, Q_t^\frac{b}{b+1}\, dt + \hat \lambda\, Q_t\, dB_t ,\]
or alternatively $dQ_t= \frac{\lambda^2}{2}\,\epsilon\, X_t^b\, dt + \lambda\, X^{b+1}_t\, dB_t $, where we have absorbed numerical multiplicative factors in redefining $\hat \epsilon$ and $\hat \lambda$: $\hat \lambda = \lambda\, (b+1)$ and $\hat \epsilon = \epsilon (b+1)^\frac{-1}{b+1}$.
What motivates this change of variable is that $Q_t$ is a local martingale when $\epsilon =0$: the function $q(x):=\frac{x^{b+1}}{b+1}$ is  scale function for the process $X_t$, see e.g. \cite{ito-mckean}.

\begin{lem}
  Consider the time change trading the standard time $t$ for the effective time $\tau$ defined by $d\tau = \lambda^2 X_t^{2(b+1)}\,dt$ so that $W_\tau$ defined via $dW_\tau = \lambda X_t^{b+1}\, dB_t$  is a Brownian motion in the time parameter $\tau$. In the limit $\epsilon\to 0$, the process $Q$ parametrized by $\tau$ is a reflected Brownian motion:
\[ Q_\tau=|\tilde W_\tau|,\]
with $dW_\tau=\mathrm{sign}(\tilde W_\tau)d\tilde W_\tau$. 
\end{lem}

{\bf Proof:} 
The proof relies on the use of the approximate Skorokhod decomposition presented above. Tanaka's equation \cite{yenyor} gives an interpretation of the Skorokhod decomposition of a Brownian sample: if $\tilde W_\tau$ is a Brownian motion, $|\tilde W_\tau|$ satisfies $d |\tilde W_\tau|=dL_\tau+ \mathrm{sign}(\tilde W_\tau)d\tilde W_\tau$ where $L_\tau$ is the local time of $\tilde W_\tau$ at the origin, and by L\'evi's charaterization theorem, $W_\tau:=\int_0^\tau  \mathrm{sign}(\tilde W_\sigma)d\tilde W_\sigma$ is a Brownian motion. Thus $|\tilde W_\tau|=L_\tau +W_\tau$, and it is easily seen that this is the Skorokhod decomposition of $W_\tau$. \\
The SDE for $Q$ parametrized with $\tau$ reads: 
\[ dQ_\tau = dL^{(\hat \epsilon)}_\tau + dW_\tau,\quad 
dL^{(\hat \epsilon)}_\tau :=\frac{\hat \epsilon}{2} Q_\tau^{-\frac{b+2}{b+1}}\, d\tau .\]
This is exactly \autoref{eq:skoapprox} with obvious changes in notation and we apply the effective version of Skorokhod's lemma to conclude that for each Brownian motion sample $W_\tau$ we have that  $L^{(\hat \epsilon)}$ converges to $L_\tau$ as $\epsilon \to 0^+$, uniformly on compact time  intervals. $\square$

This lemma gave the expression of $X$ in terms of the Brownian motion $\tilde W$. It remains to find the relation between $\tau$ and the original physical time $t$.

\begin{lem}
  In the scaling limit $\epsilon\to0$, $\lambda\to\infty$ with $J= \lambda^2\epsilon^{b+1}$ fixed, one has:
\beq \label{LtoT}
 L_\tau = \frac{J}{2\Gamma(b+1)}\, t.
 \eeq
 with $L_\tau$ the $\tilde W_\tau$ local time at the origin.
\end{lem}

{\bf Proof:}
The expression of $L^{(\hat \epsilon)}_\tau$ in terms of the original time is via $dL^{(\hat \epsilon)}_\tau =\frac{\hat \lambda^2}{2}\, \hat \epsilon\, Q_t^{b/b+1}\, dt=\frac{\lambda^2}{2}\,\epsilon\, X_t^b\, dt$, so that the relation between $L^{(\hat \epsilon)}_\tau$ and $t$ can be written in an integral form:
\[ L^{(\hat \epsilon)}_\tau= \frac{\lambda^2\epsilon}{2}\, \int_0^t ds\, X_s^b = \frac{J}{2}\, \int_0^t ds\, \Big(\frac{X_s}{\epsilon}\Big)^b  .\]
Let us change time and variable and set $Y_{s}:= \frac{X_t}{\epsilon}$ with $s=\lambda^2 t$. The relation between $L^{(\hat \epsilon)}_\tau$ and $t$ can then be written as
\[ L^{(\hat \epsilon)}_\tau = \frac{J}{2}\, \int_0^t ds\, \Big(\frac{X_s}{\epsilon}\Big)^b = \frac{Jt}{2}\, \frac{1}{\lambda^2 t}\int_0^{\lambda^2 t} \hskip -0.3 truecm ds\, Y_s^b.\]
The process $Y_s$ solves the SDE $dY_s= \frac{1}{2}(1-bY_s)ds+ Y_s d\tilde B_s$ in which $\epsilon$ and $\lambda$ do not appear. We can thus apply the ergodic theorem to get
 \[ \lim_{\lambda\to \infty}  \frac{1}{\lambda^2 t}\int_0^{\lambda^2 t} \hskip -0.3 truecm ds\, Y_s^b = \langle Y^b \rangle_\mathrm{inv}.\]
 The invariant measure for $Y$ is that of $X/\epsilon$ so that
 \[ \langle Y^b \rangle_\mathrm{inv} = \frac{1}{\Gamma(b+1)} \int \frac{dy}{y^{b+2}} e^{-1/y} y^b =  \frac{1}{\Gamma(b+1)}.\]
This yields $2 \Gamma(b+1)\, L_\tau= {Jt}$ as announced. $\square$

To summarize, we have proven that in the scaling limit $X_t^{b+1}= (b+1) |\tilde W_\tau|$ (equality in law) with $\tilde W_\tau$ is a normalized Brownian motion (with respect to $\tau$) and that the relation between $\tau$ and $t$ is through the $\tilde W_\tau$ local time $L_\tau$ at the origin.

The law of the trajectories $X_t$ in the scaling limit follows from classical results on excursions. The set of $\tau$s such that $|\tilde W_\tau|>0$ is open in $\mathbb{R}^+$ and as such it is the union of a countable number of disjoint open intervals. On each of these intervals, the local time $L$ of $|\tilde W|$ is constant but $|\tilde W|$ makes an excursion. Thus when the local time is used as a parameter, each excursion becomes instantaneous and draws a vertical segment, a spike, whose height is the maximum of the excursion. It is a classical result (see e.g. \cite{yenyor}) that, when parameterized by $L$, the maxima of the excursions of $|\tilde W|$ form a Poisson point process on $\mathbb{R}^+\times \mathbb{R}^+$ with intensity $dL \frac{dm}{m^2}$. Thus in the scaling limit the trajectories of $X$ are made of spikes whose tips form a Poisson point process on $\mathbb{R}^+\times \mathbb{R}^+$ with intensity $\frac{(b+1)^2J}{\Gamma(b+1)} dt \frac{dm}{m^{2+b}}$. From this correspondence, more features of $X$ in the scaling limit can be computed, and in particular anomalous behaviors. We refer to ref.\cite{BBTskospikes}, which treats the case $b=0$ in detail, for more details.

\subsection{Scaling limits via passage times} \label{ssec:slpt}

There is another useful approach to study the scaling limit, which is based on standard tools in the study of diffusions in one dimension \cite{ito-mckean}.
Let us quote once again our starting point \autoref{sde:b0} that we rewrite here:
\beq \label{sde:b0ter}
dX_t = \frac{\lambda^2}{2}\,(\epsilon - b X_t)dt + \lambda\, X_t\, dB_t .
\eeq 
We view this equation as defining a family of probability measures $\mathbb P_{\lambda,\epsilon}$ on some canonical space of functions (say continuous functions on $\mathbb{R}^+$). 

We are interested in a non-trivial limit for $\mathbb P_{\lambda,\epsilon}$ with $\epsilon$ carefully adjusted to go to $0$ at an appropriate rate when $\lambda \to \infty$. The role played by $\epsilon$ in the above equation is noticeable only when $X_t$ is small. Thus we have the following rules of thumb. Suppose $A$ is a property (i.e. a measurable set) of trajectories \\ 
-- If $A$ can be decided without ever looking at times when $X$ is small, then the limits of large $\lambda$ and small $\epsilon$ can be taken independently, i.e. $\lim_{\text{scaling}} \mathbb P_{\lambda,\epsilon}(A)=\lim_{\lambda \to \infty,\epsilon \to 0^+}\mathbb P_{\lambda,\epsilon}(A)$.\\
-- If $A$ cannot be decided without ever looking at times when $X$ is small, be careful!

Let us illustrate this with three computations.

\underline{Exit probabilities:} Let $0 \leq x < y < z$. Let $P_{[x,z]}(y)$ denote the probability that the process started at $y$ exits $[x,z]$ at $z$. The Markov property implies that $P_{[x,z]}(X_t)$ is a local martingale so that by It\^o's formula $\frac{\lambda^2}{2} \left( (\epsilon - by)  P'_{[x,z]}(y)+ y^2P''_{[x,z]}(y)\right)=0$ (where $'$ denotes the derivative with respect to $y$) with boundary conditions $P_{[x,z]}(x)=0$ and $P_{[x,z]}(z)=1$. As expected, $\lambda$, which is just an overall time scale, factors out of this time-independent quantity. So in this case, it it trivial to see that taking the scaling limit or taking independently the limits of large $\lambda$ and small $\epsilon$ do yield the same outcome. This is consistent with our rule of thumb (which applies if $x >0$ because in that case indeed, we can decide if the exit is at $z$ without ever looking at trajectories going below $x$, i.e. without ever looking at trajectories coming close to regions where the $\epsilon$ term is important). The fact that it is valid even if $x=0$ is not predicted by the rule of thumb, and in fact there is some kind of revenge, see below. 

The equation for $P_{[x,z]}(y)$ is easily solved to yield
\[ P_{[x,z]}(y)=\frac{\int_x^y u^b e^{\epsilon/u}du }{\int_x^z u^b e^{\epsilon/u} du}.\]

Note that, $\forall \epsilon >0$, 
$P_{[0,z]}(y)=\lim_{x \to 0^+} P_{[x,z]}(y)=1$, implying that the process started at $y >0$ never visits $0$ but visits arbitrary large values $z$ with probability $1$. On the other hand, by a trivial application of dominated convergence say, for $x >0$, $\lim_{\epsilon \to 0^+} P_{[x,z]}(y)=\frac{y^{b+1}-x^{b+1}}{z^{b+1}-x^{b+1}}$ which is indeed the $\epsilon=0$ result, but whose limit as $x \to 0^+$ is $\frac{y^{b+1}}{z^{b+1}}\neq 1$, i.e. the limits $\epsilon \to 0^+$ and $x \to 0^+$ do not commute. The point is that when $\epsilon=0$ the trajectories started at $y >0$ converge to $0$ and have a nonzero probability to remain below $z$ for all times whenever $z >y$. 

So even for this very simple quantity, some subtleties are present. 

\vspace{.3cm}

\underline{Down-crossing times:} For $0 < x < y$, let $T_{y\to x} $ be the time it takes to go from $y$ to $x$. We call this a down-crossing because $x < y$. Our rule of thumb suggests that the law of this time (which is clearly a stopping time) is not sensitive to the way the limits of large $\lambda$ and small $\epsilon$ are taken, because the part of a trajectory needed to compute its $T_{y\to x} $ never goes below $x>0$ fixed. 
We concentrate on the Laplace transform $\mathbb{E}[e^{-\sigma T_{y\to x} } ]$. 

We can write:
\[ \mathbb{E}[e^{-\sigma T_{y\to x} }] = e^{-\int_x^y \phi(u,\sigma)du}= \frac{\psi(y,\sigma)}{\psi(x,\sigma)},\]
for some appropriate functions $\phi$ and $\psi$. These formul\ae\ express that if $u \in [x,y]$, to go from $y$ to $x$ you have to go from $y$ to $u$ and then from $u$ to $x$, so the $T_{y\to x}$ is the sum $T_{y\to u} + T_{u\to x}$ of two random variables which are independent by the strong Markov property. 

Furthermore the process $e^{-\sigma t} \psi(X_t,\sigma)$ is a local martingale, so that by It\^o's formula $\psi$ satisfies a second order differential equation, 
\[ u^2\psi''(u)+ (\epsilon-bu)\psi'(u)= \frac{2\sigma}{\lambda^2}\psi(u),\]
and $\phi$ a Ricatti type equation, namely 
\[u^2 ( -\phi'(u) + \phi^2(u)) - (\epsilon - b u)\phi(u) = 2\sigma/\lambda^2.\]
Because we are looking at a down-crossing, the boundary conditions for $\phi$ are at $u\to +\infty$ where the $\epsilon$ term is negligible. So indeed the $\epsilon \to 0^+$ and $\lambda \to \infty $ can be taken independently. In the case at hand the solution at $\epsilon=0$ can be computed explicitly. One finds $\phi(u)=\frac{\sqrt{(b+1)^2+8\sigma/\lambda^2}-(b+1)}{2u}$ i.e. 
\[ \lim_{\epsilon \to 0^+}\mathbb{E}[e^{-\sigma T_{y\to x} }] = \left(\frac{x}{y}\right)^{\frac{\sqrt{(b+1)^2+8\sigma/\lambda^2}-(b+1)}{2}}\]
The  corrections in $\epsilon$ are straightforward to compute, especially because the aforementioned scale invariance of \autoref{sde:b0ter} translates into the fact that $u\phi(u)$ is a function of the variable $u/\epsilon$. Order by order in perturbation theory one gets an inhomogeneous linear first order differential equation with only one solution compatible with the boundary condition at large $u$ so that $u\phi(u)\sim \sum_n c_n \left(\frac{\epsilon}{u}\right)^{n}$ where  $c_0:=\frac{\sqrt{(b+1)^2+8\sigma/\lambda^2}-(b+1)}{2}$ and the $c_n$s are, like $c_0$, explicit functions of $b$ and $\sigma \lambda^{-2}$ that can be computed mechanically at least for the first few. We conclude:
\begin{lem}
When $\epsilon \to 0^+$ and $\lambda \to \infty $ independently, and in particular in the scaling limit when $\lambda^2 \epsilon^{b+1}=:J$ is kept fixed, the random variable $T_{y\to x} $ for $0 < x < y$ scales like $\lambda^{-2}$ and in particular its law converges to that of the zero random variable. 
\end{lem}
In words, when $\lambda^{-2}$ and $\epsilon \to 0^+$ it takes no time to go from $y$ to $x$, more precisely a time of order $\lambda^{-2}$. 

During a down-crossing from $y$ to $x$ , the process has a maximum $M_{y \to x}$ and the same kind of argument shows that the joint law of $(T_{y\to x},M_{y \to x})$  has a limit when $\epsilon \to 0^+$ and $\lambda \to \infty $ independently. Even if the down-crossing time and the maximum are not independent, they become so at large $\lambda$ and the limiting law can be guessed from the two computations above: by the first computation the probability  that $M_{y \to x}>m$ is $P_{[x,m]}(y)$ whose limit is $\frac{y^{b+1}-x^{b+1}}{m^{b+1}-x^{b+1}}$, and conditionally on $M_{y \to x}$ the time is takes to do the down-crossing scales like $\lambda^{-2}$ and in particular its law converges to that of the zero random variable. We shall not give the details because this is intuitive given the second computation. Let us note however that making this rigorous would not be difficult: we could describe the down-crossing from $y$ to $x$ with maximum $m$ as the concatenation of two independent processes, the first is $X$ started at $y$ and conditioned to reach $m$ before $x$, and the second is $X$ stated at $m$ and conditioned to reach $x$ before touching $m$ again. Both pieces are described as diffusions via Girsanov's theorem, the additional drift term due to conditioning being explicit in terms of $P_{[x,m]}(y)$. For both pieces the trajectories do not go below $x$ so that the limit $\epsilon \to 0^+$ can be taken straightforwardly and independently of the large $\lambda$ limit.

\vspace{.3cm}

\underline{Up-crossing times:} For $0 < y < z$, let $T_{y\to z} $ be the time it takes to go from $y$ to $z$. We call this an up-crossing because $z >  y$. Our rule of thumb says that we should be careful, because for the law of $T_{y\to z}$ the trajectories that go close to $0$ might make a contribution. And sure they do: for $\epsilon=0$ the trajectories converge to $0$ and do not always reach $z$. For small $\epsilon$, they either reach $z$ quite early or they start to converge to $0$ and only the presence of the $\epsilon$ term gives them a kick and a new chance to go up to $z$. We know from the computation of exit probabilities that, due to these kicks, at some point $z$ will be reached. But the smaller $\epsilon$, the larger the number of needed kicks. This is the basis of the intuitive argument given in the introduction and which quantifies the careful rescaling of time by a factor $\lambda$ depending on  $\epsilon$ needed to reach $z$ in finite time. We shall now give a rigorous treatment. 

\begin{lem}
 
In the scaling limit $ \lambda\to \infty,\ \epsilon\to 0$ with $\lambda^2 \epsilon^{b+1}=:J$ fixed, one has:
\beq \label{dist:T}
\lim_{\text{scaling}} \mathbb{E}[e^{-\sigma T_{y\to z} } ] = \frac{ 1 + \sigma\, \hat J^{-1}\, q(y)}{ 1 + \sigma\, \hat J^{-1}\, q(z)} ,
 \eeq 
with $q(x)= x^{b+1}/(b+1)$ and $\hat J = J/2\Gamma(b+1)$.
\end{lem}

{\bf Proof:}
The arguments used for down-crossings can be repeated to show that one may write 
\[ \mathbb{E}[e^{-\sigma T_{y\to z} } ] = e^{-\int_y^z \phi(u,\sigma)du}= \frac{\psi(y,\sigma)}{\psi(z,\sigma)},\]
and the functions $\psi$ or $\phi$ satisfy differential equations.
For the SDE (\ref{sde:b0}), it reads 
\[ u^2\psi''(u)+ (\epsilon-bu)\psi'(u)= \frac{2\sigma}{\lambda^2}\psi(u)\] or equivalently 
\[u^2 ( \phi'(u) + \phi^2(u)) + (\epsilon - b u)\phi(u) = 2\sigma/\lambda^2.\] 
Beware that even if the equations for $\psi$ are the same for up- and down-crossings, the solutions are not: this time boundary conditions should be taken at small $u$. \\
The boundary condition at $u=0^+$ is straightforward: $\phi(0)=\frac{2\sigma}{\lambda^2\epsilon}$. This can be seen either by taking brutally $u=0$ in the equation for $\phi$ or by realizing that at small $u$ the equation for $X_t$ becomes deterministic and reduces to $dX_t=\frac{\lambda^2}{2}\epsilon dt$.\\
On the other hand, in the scaling limit at fixed $u$  one expects that the terms involving $\epsilon$ and $\lambda^{-2}$ can be neglected so that the equation for $\phi$ reduces to $u^2 ( \phi'(u) + \phi^2(u)) - b u \phi(u)=0$  whose solution is $\phi_\mathrm{out}(u):=\frac{(b+1) u^b}{c + u^{b+1}}$ with $c$ an integration constant.
The "out" subscript was introduced to emphasize that $\phi_\mathrm{out}(u)$ is a kind of outer solution, which is expected to be good for every $u$ in the scaling limit, but could be poor at small $u$ before the scaling limit is taken. Indeed, if $b >0$ the small $u$ behavior of $\phi_\mathrm{out}$ is incompatible with the boundary condition at $u=0^+$: the true solution $\phi(u)$ changes quickly in a boundary layer to interpolate between the boundary condition at $0^+$ and the behavior of $\phi_\mathrm{out}(u)$. \\
One possibility would be to compute an inner approximation $\phi_\mathrm{in}(u)$ and do matching. While this is certainly doable, we prefer a direct analysis which gives in one stroke the scaling relation and the right boundary condition. \\
To achieve this, we rewrite the equation for $\phi$ as an integral equation. Guided by the method of variation of constants for the linearized version (where the $\phi^2$ term is neglected), we arrive at the following formula, whose correctness can of course be checked immediately by differentiation: for $y, z \in ]0,+\infty [$
\[\phi(z) \frac{e^{-\epsilon/z}}{z^b}-\phi(y) \frac{e^{-\epsilon/y}}{y^b} + \int_y^z du \, \phi(u)^2 \  \frac{e^{-\epsilon/u}}{u^b} =\frac{2\sigma}{\lambda^2} \int_y^z du \,\frac{e^{-\epsilon/u}}{u^{b+2}}. \]
As long as $\epsilon >0$, when $X_t$ is small, the SDE is well-approximated by the deterministic equation $dX_t = \frac{\lambda^2}{2}\epsilon \,  dt$ which leads to the small $x$ behavior $\phi(x)\sim \frac{2\sigma}{\lambda^2\epsilon }$. Due to the crushing factor $e^{-\epsilon/y}$, the term $\phi(y) \frac{e^{-\epsilon/y}}{y^b}$  goes to as $y \to 0^+$. Thus:
\[\phi(z) \frac{e^{-\epsilon/z}}{z^b}+ \int_0^z du \, \phi(u)^2 \ \frac{e^{-\epsilon/u}}{u^b} =\frac{2\sigma}{\lambda^2} \int_0^z du \,\frac{e^{-\epsilon/u}}{u^{b+2}}. \]
Setting  $Z_{\epsilon}:=\int_0^{+\infty} du \,\frac{e^{-\epsilon/u}}{u^{b+2}}$ and denoting by 
$P_\mathrm{inv}(\epsilon,[0,z]):= Z_{\epsilon}^{-1} \int_0^z du \,\frac{e^{-\epsilon/u}}{u^{b+2}}$ the invariant measure of the interval $[0,z]$, the formula reads:
\[ \phi(z) \frac{e^{-\epsilon/z}}{z^b}+ \int_0^z du \, \phi(u)^2 \ \frac{e^{-\epsilon/u}}{u^b} =\frac{2\sigma Z_{\epsilon}}{\lambda^2} P_\mathrm{inv}(\epsilon,[0,z]).\]
The expectation of the passage time $T_{y\to z}$ is related to the first order in the expansion of $\phi(z)$ in powers of $\sigma$, i.e. to the linearization of the above equation. Quantitatively
\[
\frac{\partial \mathbb{E}[T_{y\to z}]}{\partial z}\frac{e^{-\epsilon/z}}{z^b}=\frac{2 Z_\epsilon}{\lambda^2}P_\mathrm{inv}(\epsilon,[0,z]).
\]
Remember our aim is to find a scaling for which $T_{y\to z}$, and in particular its expectation, remains finite when $\epsilon \to 0^+$. As $\lim_{\epsilon \to 0^+}P_\mathrm{inv}(\epsilon,[0,z])=1^-$ for each $z>0$, the only way this can occur is by adjusting $Z_\epsilon \lambda^{-2}$ to have a finite limit. Thus we define the scaling limit as $\lambda\to\infty$, $\epsilon\to 0^+$, with $\lambda^2/Z_\epsilon=2\hat J$ fixed. This scaling relation is of course identical to the one obtained above via the study of the Skorokhod decomposition. Now if in the three terms equation for $\phi$ two terms have a limit when $\epsilon \to 0^+$, namely $\phi(z)$ and $\frac{2\sigma Z_\epsilon}{\lambda^2}P_\mathrm{inv}(\epsilon,[0,z])$, the third term, $\int_0^z du \, \phi(u)^2 \frac{e^{-\epsilon/u}}{u^b}$ must have a limit too, and the only question is whether the integral of the limit is the limit of the integral. The only possible problem is near the origin, within  the boundary layer. Replacing  
$\phi(u)$ by $\frac{2\sigma}{\lambda^2\epsilon}$ in a layer of thickness $\sim \epsilon$, the contribution to the integral is seen to be of order $O(\lambda^{-4} \epsilon^{-(b+1)})$ which is very small compared to $\lambda^{-2} \epsilon^{-(b+1)}$ at large $\lambda$. Thus there is no anomalous contribution due to the boundary layer, and at $\epsilon=0$ we obtain the equation:
\[ \phi(z) z^{-b}+ \int_0^z du \, \phi(u)^2 u^{-b}  =\frac{\sigma}{\hat J}.\]
The solution is easily found to be\footnote{Note that this makes  $\int_0^z du \, \phi(u)^2 u^{-b}$ finite because $b > -1$.}:
\[ \phi(z)=\frac{z^b}{\frac{\hat J}{\sigma}+\frac{z^{b+1}}{b+1}}, \]
from which the announced formula for $\lim_{\text{scaling}} \mathbb{E}[e^{-\sigma T_{y\to z} } ]$ follows. $\square$

A salient feature that will be present in the more complicated cases as well is that the generating function (\ref{dist:T}) for the passage times is a ratio ${\psi(y,\sigma)}/{\psi(z,\sigma)}$ with $\psi(x,\sigma)$ a  polynomial of degree one in $\sigma$. The Laplace transform can be inverted easily to yield: 

\begin{cor}
	In the scaling limit $ \lambda\to \infty,\ \epsilon\to 0$ with $\lambda^2 \epsilon^{b+1}=:J$ fixed, the law of $T_{y\to z}$ converges to a mixture of $\delta_0$ (the Dirac point measure at the origin) with weight $\frac{y^{b+1}}{z^{b+1}}$ and an exponential distribution of parameter $\hat J\frac{b+1}{z^{b+1}}$ with weight $1-\frac{y^{b+1}}{z^{b+1}}$.
\end{cor}

This result on down-crossings alone, possibly appealing to Okam's razor, contains all the ingredients of our previous results on exit probabilities and up-crossings. To wit:\\
-- The weight of $\delta_0$ is simply the probability, starting from $y$, to reach $z>y$ before $0$ when $\epsilon=0$ (but keep in mind the caveat about non commuting limits)\\
-- The weight for the exponential distribution is the probability, starting from $y$, to reach $0$ before $z>y$ when $\epsilon=0$. 

So the simplest interpretation is that in the scaling limit either $z$ or $0$ is reached in no time. As $z$ can be arbitrary large, $0$ must be reached in no time anyway, possibly after after having visited $z$. And then, once at $0$, the process waits an exponential time to reach $z$ again.  Crucial for this last point is that the parameter for the exponential distribution depends only on the endpoint $z$, not on $y$. 

\begin{figure}
	\begin{center} \includegraphics[width=0.90\textwidth]{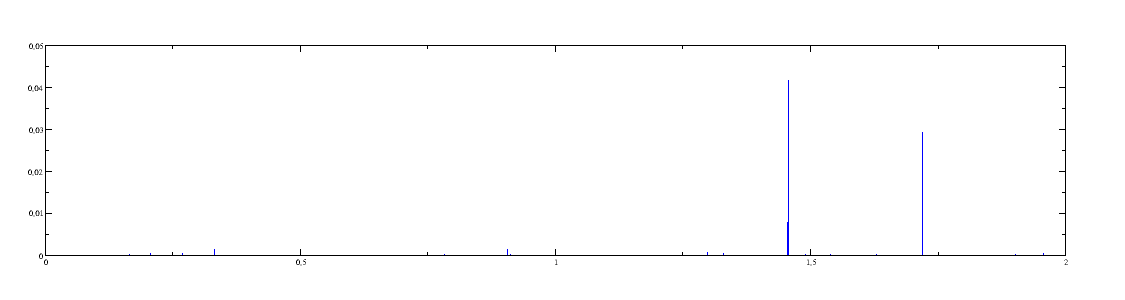} \end{center} 
	
	\vspace{-1.4cm}

	\begin{center} \includegraphics[width=0.90\textwidth]{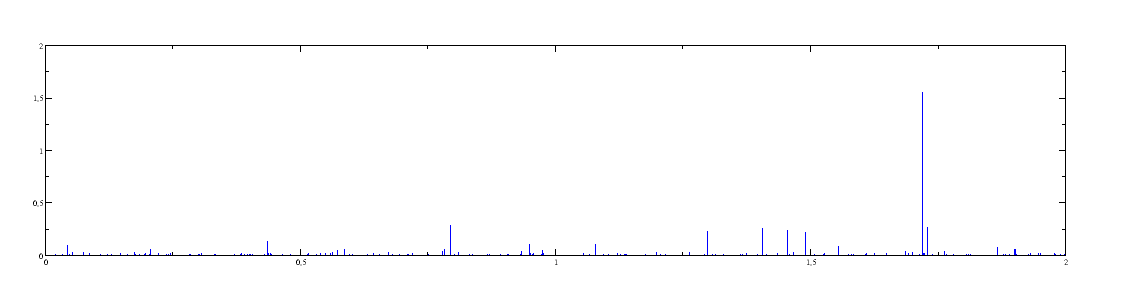} \end{center} 
	
	\vspace{-1.4cm}
	
	\begin{center} \includegraphics[width=0.90\textwidth]{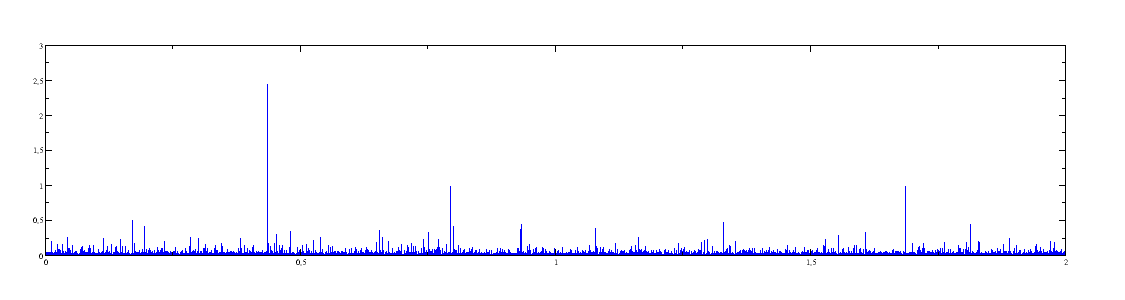} \end{center} 
	
	\vspace{-1.4cm}
	
	\begin{center} \includegraphics[width=0.90\textwidth]{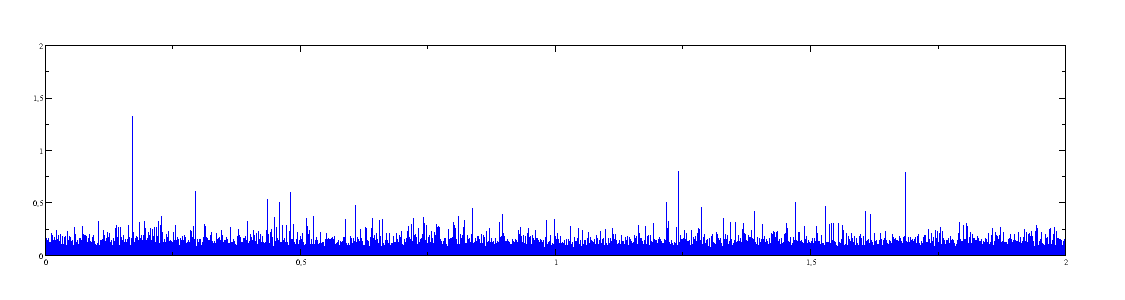} \end{center} 
	
	\vspace{-1.4cm}
	
	\begin{center} \includegraphics[width=0.90\textwidth]{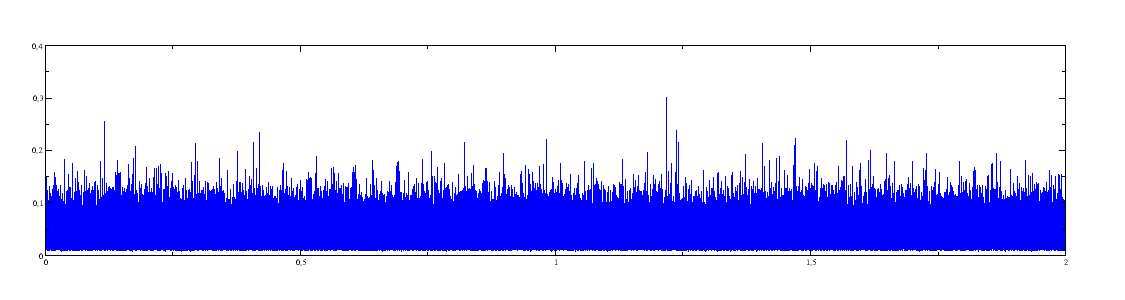} \end{center} 
	 
	\caption{\emph{Samples with $b=-1/2,b=0,b=1,b=3,b=10$, with large $\lambda$ but $\lambda^2\epsilon^{b+1}=1$. The tips of the spikes are approximate samples of a Poisson point process with a measure proportional to $x^{-(b+2)} dt dx$. For large but fixed $\lambda$, the quality of the approximation decreases with larger $b$s.}}
	\label{fig:linear_spikes}
\end{figure}

We illustrate our results with \autoref{fig:linear_spikes} for different values of $b$.

\subsection{The point process of maxima and its scaling limit} \label{ssec:ppmsl}

The spiky features of the samples in the scaling limit, and the results obtained above, suggest that the following approach is relevant. 

Fix $\gamma$ and $\varepsilon$. Let $0 < \delta_- < \delta_+$ be two numbers which are arbitrary for the time being. We shall let them go to $0$ later on. Take $X_0=\delta_-$ and define the crossing times $U_k,k=0,1,2,\cdots$ by $U_0:=0$, $U_1:=\inf\{0 < t, X_t=\delta_+\}$, $U_2:=\inf\{U_1 < t, X_t=\delta_-\}$, $U_3:=\inf\{U_2 < t, X_t=\delta_+\}$ and so on. Between $U_{2k}$ and $U_{2k+1}$ , $k=0,1,\cdots$ the process $X_t$ makes an up-crossing from $\delta_-$ to $\delta_+$ while remaining below $\delta_+$. Between  $U_{2k-1}$ and $U_{2k}$ $k=1,2,\cdots$ the process $X_t$ makes a down-crossing from $\delta_+$ to $\delta_-$ while remaining above $\delta_-$. 

Let $M_k:=\sup \{X_t, t\in [U_{2k-1},U_{2k}]\}$. Let $PP_{\delta_-,\delta_+}:=\{(U_{2k},M_k)\in \mathbb{R}^+\times \mathbb{R}^+\}_{k=1,2,\cdots}$. We call $PP_{\delta_-,\delta_+}$ the point process of maxima. It is a random set of points in $\mathbb{R}^+\times \mathbb{R}^+$. We note an important restriction property : if $x \geq \delta_+$ then $PP_{\delta_-,x}=PP_{\delta_-,\delta_+}\cap \mathbb{R}^+\times [x,+\infty[$, a plain consequence of the continuity of trajectories. 

If $B$ is any (say Borel) subset of $]0,+\infty[$ bounded away from $0$, we define the counting process $N_t(B)$, $t \in \mathbb{R}^+$ which is the number of entrances of $PP_{\delta_-,\delta_+}$ in $B$ before $t$, in mathematical notation $N_t(B):=\# \{k \geq 1, \, U_{2k} \leq t\text{ and } M_k \in B \}$. Let $k_1 < k_2\cdots$ be the ordered sequence of $k$s such that $M_k \in B$, and set $U_{n}(B):=U_{2k_n}$ , $M_n(B):=M_{k_n}$. Then the $U_{n}(B)$s, $n=1,2,\cdots$ are the jump times of $N_t(B)$. For $B=[\delta_+,+\infty[$, $M_n(B):=M_{n}$ and $U_{n}(B)=U_{2n}$.

We shall now proceed to show that in the scaling limit the point process of maxima converges weakly (i.e. in law) to a Poisson point process. 

Define $T_k:=U_k-U_{k-1},k=1,2,\cdots$.  By the strong Markov property the two sequences $T_{2k+1}, k=0,1,\cdots$ and $(T_{2k},M_k)$, $k=1,2,\cdots$ are independent of each other and each individual sequence is $iid$, though $T_{2k}$ and $M_k$ are not independent. In particular, the sequence $(T_{2k-1}+T_{2k},M_k)$, $k=1,2,\cdots$ is $iid$, which may be rephrased by saying the the sequence $(U_{2k},M_k)$ is a renewal process. For the same reasons, for any Borel subset $B$ of $\mathbb{R}^+$, the sequence $(U_{n}(B),M_n(B))$ is a renewal process. 

We start by computing, in the scaling limit and for any $z \geq \delta_+ >0$ the law of $(U_1(B),M_1(B))$ in $PP_{\delta_-,\delta_+}$ when $B=[z,+\infty[$. By restriction, the entrances in $B$  for $PP_{\delta_-,\delta_+}$ and for $PP_{\delta_-,z}$ are the same, so we take  $\delta_+=z$ for this computation. Started at $\delta_-$ at $t=0$, the trajectory reaches $z$ at $U_1=T_1=T_{\delta_-\to z}$, and reaches a maximum $M_1$ during the down crossing from $z$ to $\delta_-$ which is achieved at $U_2$. We need the law of $(U_1(B),M_1(B))=(U_1,M_1)=(T_1+T_2,M_1)$ in the scaling limit. But $T_1=T_{\delta_-\to z}$ and $(T_2,M_1)$  are independent and their law in the scaling limit where computed in the previous Section. We conclude

\begin{lem}
	In the scaling limit, the entrances in $[z,+\infty[$ of the point process $PP_{\delta_-,\delta_+}$ ($z > \delta_+$) form a renewal process where the waiting times $T$ and the entrance point $M$ are independent and have distribution 
	\[\mathbb{P}_{\text{scaling}}(T\geq t)=\left(\frac{\delta_-} {z}\right)^{b+1}+\left(1-\left(\frac{\delta_-} {z}\right)^{b+1} \right)e^{-\hat J\frac{b+1}{z^{b+1}}t},\; t \geq 0 \] and
	\[ \mathbb{P}_{\text{scaling}}(M \geq m) = \left(\frac{z}{m}\right)^{b+1}, \; m\geq z.\] 
	In particular, taking the scaling limit and then letting $\delta_+, \delta_- \to 0^+$, for any $z >0$, the counting process $N_t([z,+\infty[)$ of entrances in $[z,+\infty[$, converges in law to a standard Poisson process of parameter $\hat J\frac{b+1}{z^{b+1}}$, and the points of entrance form an independent $iid$ sequence with distribution density $(b+1)z^{b+1}\frac{dm}{m^{b+2}}$.
\end{lem}

If $B$ is any Borel subset of $\mathbb{R}^+$ set $\mu(B):=\int_B \frac{(b+1)dm}{m^{b+2}}$, which is $< +\infty$ if $B$ is bounded away from $0$. A standard argument shows that:

\begin{prop}
	For any Borel subset $B$ of $\mathbb{R}^+$ bounded away from $0$, taking the scaling limit and then letting $\delta_+, \delta_- \to 0^+$, the counting process $N_t(B)$ of entrances in $B$, converges in law to a standard Poisson process of parameter $\hat J\mu(B)t$, and the points of entrance form an independent $iid$ sequence with distribution $\frac{\mu(.)}{\mu(B)}.$
\end{prop}

A Poisson point process with intensity measure $\nu$ on a measure space $(U,\mathcal E)$ (satisfying some supplementary technical assumptions) is defined as a collection of random variables $N(A)$ indexed by the measurable subsets $A\in \mathcal E$ such that the $N(A)$s for families of disjoint $A$s are independent, and if $\nu(A) < +\infty$, $N(A)$ is a Poisson random variable with parameter $\nu(A)$. When $E$ is a product measurable space $E=\mathbb{R}^+ \times F$ and $\nu$ is the product of the Lebesgue measure on $\mathbb{R}^+$ and a measure $\mu$ on $F$, the Poisson point process with intensity measure $\nu$ has an alternative characterization: for each measurable subset $B$ of $F$ with $\mu(B) < +\infty$, the counting process $N_t(B)$ of entrances in $B$ is a standard Poisson process on $\mathbb{R}^+$ with parameter $\nu(B)$ and the points of entrance in $B$ form an independent $iid$ sequence with distribution $\frac{\mu(.)}{\mu(B)}.$ This follows straightforwardly from the particular case when $F$ is a singleton. Thus we have proved:

\begin{prop}
	Taking the scaling limit and then letting $\delta_+, \delta_- \to 0^+$, the point process of maxima converges in law to a time homogeneous Poisson point process with intensity $d\nu=\hat J dt \frac{(b+1)dm}{m^{b+2}}$.
\end{prop}

We are now able to prove one of our main results, which implies that in the scaling limit the trajectories can be reconstructed with arbitrary precision from the process of maxima $PP_{\delta_-\delta_+}$. 

Define $K_{\delta_-,\delta_+}:=\mathbb{R}^+\times [0,\delta+] \bigcup \cup_{k=1,2,\cdots}  ([U_{2k-1},U_{2k}]\times [0,M_k])$. It is the random set of points that are below $\delta_+$ on up-crossings and below the maximum on down-crossings. In particular, $K_{\delta_-,\delta_+}\supset \{ (t,X_t)\}_{t\in\mathbb{R}^+}$ so it gives a rough approximation of the trajectory as a set. Fix an arbitrary $\Delta>0$ and define $K^{\Delta}_{\delta_-,\delta_+}:= \mathbb{R}^+ \times [0,\delta_+] \bigcup \cup_{k=1,2,\cdots} [U_{2k}-\Delta, U_{2k}] \times [0,M_k]$. Observe that $K^{\Delta}_{\delta_-,\delta_+}$ is uniquely determined by $PP_{\delta_-\delta_+}$ and is a collection of thickened spikes of thickness $\Delta$ emerging from a "background" of height $\delta_+$, i.e. $K^{\Delta}_{\delta_-,\delta_+}$ is a spiky object in the limit of small $\Delta$ and $\delta_+$. 

From the above discussion, we know that for every $t\in \mathbb{R}^+$, for all $\lambda,\epsilon$ and also in the scaling limit, there is only a finite number of points in $PP_{\delta_-\delta_+}$ with time coordinate $\leq t$. In the scaling limit, and when $\delta_- \to 0^+$ this number is a Poisson random variable with parameter $\hat J\frac{b+1}{\delta_+^{b+1}}t$. And from the convergence result for the joint law of $(T_{2k},  M_k)$ we know that the $T_{2k}$ are independent and small in the scaling limit: more precisely, for any $\Delta, t >0$, with probability going to $1$ in the scaling limit, all down-crossings from $\delta_+$ to $\delta_-$ before time $t$ will take a time less than $\Delta$. Thus we have proved:

\begin{prop}
	For every choice of $t>0$,  $\Delta$ and $\delta + > \delta_- >0$  the probability that, up to time $t$, $K_{\delta_-,\delta_+}$ (and in particular the graph of the trajectory) is contained in  $K^{\Delta}_{\delta_-,\delta_+}$ goes to $1$ in the scaling limit. 
\end{prop}

In the above proposition, the interesting domain for the parameters is $t$ large and  $\Delta$, $\delta +$, $\delta_-$ small.

Combined together, the propositions giving, in the scaling limit, the law of the point process of maxima and the inclusion of the graph of trajectories in a spiky set defined solely in terms of the point process of maxima, provide a complete description of trajectories in the scaling limit.

\section{Strong noise limit for a more general class of SDEs}
\label{sec:more}

We generalize the previous proofs to the more general SDEs considered in \autoref{prop:moregeneral} (cf. \autoref{sec:intro}):
\beq\label{sde:bqn}
 dX_t = \frac{\lambda^2}{2}\,(\epsilon X_t^q - b X_t^n)dt + \lambda\, X_t^k\, dB_t,
 \eeq
with $q,\ n$ and $k$ positive numbers (not necessarily integers) satisfying $n=2k-1>q\geq 0$ and $b >-1$. We look for the large noise limit $\lambda\to\infty$, $\epsilon\to 0$, with the scaling relation $\lambda^2\, \epsilon^\frac{b+n}{n-q}=:J$ fixed. The proofs are essentially identical to those of previous Section (which correspond to $q=0$, $n=k=1$) so that we shall be brief in our presentation. 

The condition $n=2k-1$ ensures good scaling properties which will be needed to complete certain proofs. In particular, we may scale out the parameters $\lambda$ and $\epsilon$ by defining $Y_s:= \epsilon^{-\frac{1}{n-q}}\, X_t$ with $s=\lambda^2\, \epsilon^\frac{k-1}{n-q}t$ which satisfies a $\lambda$ and $\epsilon$ independent SDE, $dY_s= \frac{1}{2}(Y_s^q-bY^n_s)ds+Y^k_sd\tilde B_s$, provided $n=2k-1$.

Another aspect of scaling is the connexion with Bessel processes for $\epsilon=0$. The case $k=1$ has already been studied and it is easily checked that for $k\neq 1$  the process $R_t:=\frac{X_t^{1-k}}{k-1}$ is, for $\epsilon=0$ and up to a rescaling of time by a factor $\lambda^2$, a $d$-dimensional Bessel process where $d=\frac{2k-1+b}{k-1}=2+\frac{b+1}{k-1}$. Thus if $k >1$ the condition $b >-1$ is a necessary and sufficient condition for $R_t\to \infty$ at large $t$, while if $\frac{1}{2} < k <1$ the condition  $b >-1$ is a necessary and sufficient condition for $R_t$ to visit $0$ with probability $1$. Going back to the $X$ variable, we conclude that $b >-1$ is a necessary and sufficient condition for $X_t$ to converge to $0$, either in finite ($\frac{1}{2} < k <1$) or infinite ($k\geq 1$) time.

We now add the simplifying assumption that $b+n >0$, which is always fulfilled if $b>0$. Then  \autoref{sde:bqn} has an invariant measure given by 
\[ P_\mathrm{inv}(x)dx = \frac{\epsilon^{-\frac{b+n}{n-q}}}{\denisZ} \frac{dx}{x^{b+n+1}}\, e^{-\frac{\epsilon }{n-q}x^{q-n}},\quad
\denisZ =(n-q)^{-\frac{b+q}{n-q}}\, \Gamma(\frac{b+n}{n-q}).\]
The conditions $n>q$ and $b+n >0$ ensure the normalizability at $0$ and $+\infty$. This measure converges to the Dirac measure at the origin $\delta_0$ in the scaling limit. 

Observe that as in the previous Section $Q_t:= \frac{X_t^{b+1}}{b+1}$ is a local martingale at $\epsilon=0$, while for general $\epsilon$,
$dQ_t = \frac{\lambda^2}{2} \epsilon\, X_t^{b+q} dt + \lambda\, X_t^{b+k} dB_t,$,
or equivalently $dQ_t = \frac{\hat \lambda^2}{2} \hat \epsilon\, Q_t^\alpha dt + \hat \lambda\, Q_t^\delta dB_t$
with $\alpha= \frac{b+q}{b+1}$ and $\delta= \frac{b+k}{b+1}$, as in \autoref{sde:autre}. We set $\hat \lambda = \lambda\, (b+1)^\frac{b+k}{b+1}$ and $\hat \epsilon = \epsilon (b+1)^\frac{q-k}{b+1}$.

To parallel the previous proofs, we change time from $t$ to $\tau$ via $d\tau= \lambda^2 X_t^{2(b+k)} dt$ and define a new Brownian motion (w.r.t. $\tau$) via $dW_\tau= \lambda\, X_t^{b+k} dB_t$. We may then rewrite the equation for $Q$ as
\[ dQ_\tau = dL^{( \epsilon)}_\tau + dW_\tau, \quad dL^{( \epsilon)}_\tau = \frac{\lambda^2}{2} \epsilon\, X_t^{b+q} dt=\epsilon\, X_t^{q-2k-b}d\tau.\]
This is exactly \autoref{eq:skoapprox} with obvious changes in notation and we apply the effective version of Skorokhod's lemma to conclude that for each Brownian motion sample $W_\tau$ the component $L^{(\epsilon)}$ of the approximate Skorokhod decomposition of $W$ converges when $\epsilon \to 0^+$ to the local time   $L_\tau$ of $W_\tau$ at $0$, uniformly on compact time intervals. In the same way $Q_\tau$ converges to $|\tilde W_\tau|$ when $\epsilon \to 0^+$.

The last step consists in relating the effective local time $L_\tau$ to the original time $t$. We again start from the relation $dL^{(\epsilon)}_\tau = \frac{\lambda^2}{2} \epsilon\, X_t^{b+q} dt$ and use the good scaling properties of \autoref{sde:bqn} to set $Y_s:= \epsilon^{-\frac{1}{n-q}}\, X_t$ with $s=\lambda^2\, \epsilon^\frac{k-1}{n-q}t$ and write
\[ L_\tau = \lim_{\lambda\to\infty} \frac{\lambda^2}{2} \epsilon\, \int_0^t X_{t'}^{b+q} dt'= \frac{Jt}{2} \lim_{T\to\infty}\frac{1}{T}\int_0^T Y_s^{b+q}ds,\]
with $T=\lambda^2 \epsilon^\frac{k-1}{n-q}$ and $J:=\lambda^2\, \epsilon^\frac{b+n}{n-q}$. We used that, by definition of the scaling limit $\lambda\to0$ and $\lambda\to\infty$, $J$ is constant. Since $Y_s$ is solution of the SDE, $dY_s= \frac{1}{2}(Y_s^q-bY^n_s)ds+Y^k_sd\tilde B_s$, we can apply the ergodic theorem to evaluate the last limit integral and get
\[ L_\tau = \frac{J}{2} \langle Y^{b+q} \rangle_\mathrm{inv} = \frac{J}{2\denisZ}\, t.\]
This concludes the discussion of \autoref{prop:moregeneral}. 

The computation of the limit distribution of the passage times associated to \autoref{sde:bqn} and the construction of the Poisson process formed by the spikes of these limit trajectories can be done as in previous Section.

\section{Strong noise limit in general}
\label{sec:general}

\subsection{Main conjectures}

We give two conjectures describing the generalization of our previous results  to a larger class of SDEs.

Though we believe that the arguments we have to support them are convincing, we chose to mention them as conjectures for two reasons. First, we have not been able to make a simple conceptual/natural list of requirements for the conclusions to apply. Second, in the general case the scaling trick that we could use in the previous examples at some crucial points in the proof to disentangle completely the roles of $\lambda$ and $\epsilon$ is not available anymore. It should be replaced by a statement that two limits commute, a fact that is ``obvious'' in the case at hand but whose precise proof has eluded us.  

Let us start again  with \autoref{sde:general} which we repeat here for convenience: 
\beq \label{eq:sde}
 dX_t = \frac{\lambda^2}{2}\big( \epsilon\, a(X_t) -  b(X_t)\big)\, dt + \lambda\, c(X_t)\, dB_t .
\eeq
We are interested in the limit $\lambda\to \infty$, $\epsilon\to 0^+$, with an appropriate scaling of $\epsilon$ vs $\lambda$. With the application to quantum mechanics in view, it is natural to see $\lambda$ as large because many measurements are performed per unit of time, and the parameter of the system ($\lambda ^2 \epsilon$) is to be tuned to ensure that the experiment is well-described by a non-trivial continuous time limit. But from a mathematical viewpoint, it is natural to start with the equation at $\lambda=1$, with conditions on the coefficients such that when $\epsilon\to 0^+$ a time scale diverges (typically the average first passage time at $z$ starting from $y <z$) and look for a change of time scale $\lambda ^{-2}$ that counterbalances the divergence. 

In the course of the argument we shall use a number of assumptions on the coefficients. Among those assumptions, some are of purely technical nature and we shall simply mention them when we use them because we have not been able to organize them in a natural way. But some assumptions have a simple conceptual interpretation and we emphasize them right now.  

\underline{Conditions when $\epsilon=0$}:\\
$i)$ The point $0$ is the only fixed point. In particular, $b(0)=0=c(0)$. \\
$ii)$ The solutions to \autoref{eq:sde} converge almost surely to $0$ (in finite or infinite time) so that the invariant measure is the point measure $\delta_0$. \\
$iii)$ There is a scale function mapping $[0,+\infty[$ to $[0,+\infty[$, i.e. the differential equation $\partial_x h_0(x)=  b(x)/2c^2(x)$ can be solved on $]0,+\infty[$ and the integral $q(y):= \int_0^y dx \, e^{2h_0(x)}$ is convergent and tends to $+\infty$ at large $x$. As usual, $h_0$ is defined only up to an additive constant so $q(y)$ is defined only up to a multiplicative constant.\\
$iv)$ The measure $\frac{dx}{c^2(x)}\, e^{-2h_0(x)}$ is an infinite (invariant, as is easily checked) measure. More precisely it is integrable on any Borel set bounded away from $0$ (and in particular at $+\infty$) but diverges at $0$.

\underline{Conditions when $\epsilon>0$, whatever small}:\\
$v)$ The behavior of solutions close to $x=0$ is governed by the $a$-term. On the other hand, we assume that the vicinity of $x=0^+$ is the only place where the $a$-term plays a significant role.\\
$vi)$ The scale functions map $]0,+\infty[$ to $]-\infty,+\infty[$, i.e. the differential equation $\partial_x h_1(x)=  -a(x)/2c^2(x)$ can be solved on $]0,+\infty[$ but the behavior of solutions is such that the primitives $\int^y dx \, e^{2(h_0(x)+\epsilon h_1(x))}=\int^y dq(x) \, e^{2\epsilon h_1(x)}$ diverge near $y=0^+$ and near $+\infty$.\\
$vii)$ The measure $\frac{dx}{c^2(x)}\, e^{-2(h_0(x)+\epsilon h_1(x))}$ is a finite (invariant, as is easily checked) measure on $]0,+\infty[$. We let $Z_\epsilon$ denote its total mass, or partition function, so that $P_\mathrm{inv}(\epsilon,[0,y]):= Z_{\epsilon}^{-1} \int_0^y \frac{dx}{c^2(x)}\, e^{-2(h_0(u)+\epsilon h_1(x))}$ is the normalized invariant measure of the interval $[0,y]$.\\
$viii)$ For $y >0$, $\lim_{\epsilon \to 0^+} P_\mathrm{inv}(\epsilon,[0,y])=1$, i.e. at small $\epsilon$ all the weight of the invariant measure is concentrated near $0$. \\
$ix)$ The two first moments of first passage times behave nicely in the small $\epsilon$ limit. 

\vspace{.2cm}

Certainly some of these conditions are vaguely stated and there are relations among them (but we have not found a way to reduce the list). Here are some general remarks.\\
-- Condition $ii)$, i.e. almost sure convergence to $0$ is clear, by a sub-martingale argument, if $b$ is a non-negative function, but our initial example shows that this is in general too strong a condition.\\
-- The scale function $q$ for the process $X_\cdot$ at $\epsilon=0$ is the ingredient to compute exit probabilities: for $0 \leq x <  y <  z$ the process started at $y$ has probability $\frac{q(y)-q(x)}{q(z)-q(x)}$ to exit $[x,z]$ at $z$. As $0$ is a fixed point, the process started at $y$ has probability $1-q(y)/q(z) >0$ to never reach altitude $z$, and by the Markov property it is seen that $iii)$  implies $ii)$.\\
-- Due to the properties we impose on $q$, we could decide to concentrate on the process $Q_t=q(X_t)$ and retrieve the general case by making the inverse change of variables in the end. Equivalently, we could content to treat the situation when the $b$-term is absent in \autoref{eq:sde}.\\
-- Condition $v)$ can manifest itself in a number of ways.  
Obvioulsy the $a$-term has to play a role important enough to counterbalance the deterministic effect of the $b$-term --and we embody this in the condition that close to $0^+$ the condition $b(x)=o(a(x))$ holds-- but also the effect of fluctuations induced by the $c$-term. \\
-- The conditions in $vi)$ are closely related to those in $v)$. The first condition in $vi)$ implies that $h_1(x)$ takes arbitrary large values close to $0^+$, and we make the simplifying assumption that $h_1(x)$ goes to $+\infty$ as $x\to 0^+$. Due to condition $iii)$, the second condition in $vi)$ will be fulfilled if $h_1(x)$ has a limit (which we are free to choose to be $0$ because $h_1$ is defined only up to an additive constant) when  $x\to +\infty$. \\
-- The first condition in $vi)$ is also a manifestation of $v)$, as it ensures that $a$-term has a repulsive effect large enough to prevent the process to reach $0^+$: for $0 \leq x <  y <  z$ the process started at $y$ has probability $\frac{\int_x^y du \, e^{2(h_0(u)+\epsilon h_1(u))}}{\int_x^z du \, e^{2(h_0(u)+\epsilon h_1(u))}}$ to exit $[x,z]$ at $z$, and this goes to $1$ when $x \to 0^+$ whatever $y$ and $z$. \\ 
-- Point $viii)$ is closely related to point $iv)$, in saying that the invariant measure for \autoref{eq:sde} behaves nicely as a function of $\epsilon$.\\
-- We shall see later that $ix)$ expresses a balance between the behavior of the scale function and of the invariant measure when $\epsilon \to 0^+$

\vspace{.2cm}

We start with:

\begin{conj}
  Let the scaling limit be defined as the limit $\lambda\to\infty$, $\epsilon\to0$ with $\frac{\lambda^2}{2Z_\epsilon}=:\hat J$ fixed. Then:\\
(i) Solutions $X_t$ of the SDE (\ref{eq:sde}) have a non-trivial limiting law in the scaling limit;\\
(ii) The limit law describes spiky trajectories $X_t$ that can be reconstructed using a reflected Brownian motion parametrized by its local time via the correspondence:
\[ |\tilde W_\tau|=q(X_t),\quad \hat J\, t = L_\tau,\]
with $L_\tau$ the local time at the origin of the standard Brownian motion $\tilde W_\tau$.
\end{conj}
-- We change variables $x\to q(x):= \int_0^xdu\,e^{2h_0(u)}$. By assumption, this function is differentiable and maps $[0,+\infty[$ bijectively to $[0,+\infty[$. As $q$ is a scale function for the process at $\epsilon=0$, the process $Q_t:=q(X_t)$ is a local martingale at $\epsilon=0$. For arbitrary $\epsilon$, $Q_t$ solves the SDE:
\[ dQ_t= \frac{\lambda^2\epsilon}{2}\, e^{2h_0(X_t)}\, a(X_t)\, dt + \lambda\, e^{2h_0(X_t)}\, c(X_t)\, dB_t.\]\\
-- We make a time change and define the effective time $\tau$ and a new Brownian motion (w.r.t. the effective time) via:
\[ dW_\tau = \lambda\, e^{2h_0(X_t)}\, c(X_t)\, dB_t,\quad d\tau = \lambda^2\, e^{4h_0(X_t)}\, c^2(X_t)\, dt .\]
Then $dQ_\tau = \epsilon\, \frac{ a(X_t)}{2c^2(X_t)}e^{-2h_0(X_t)}\, d\tau + dW_\tau =-\frac{dh_1}{dq}(Q_\tau)d\tau + dW_\tau$.  The second expression emphasizes that the local martingales associated to the diffusions $X_t$ and $Q_\tau$ via scale functions are the same (as they should be), and in particular $Q_\tau$ stays away from the origin. Letting $dL^{(\epsilon)}_\tau := \epsilon\, \frac{ a(X_t)}{2c^2(X_t)}e^{-2h_0(X_t)}\, d\tau$ denote the bounded variation part of $dQ_\tau$ we write  
\[ dQ_\tau = dL_\tau^{(\epsilon)} + dW_\tau. \]
Note that for each solution $Q_\tau$ of this equation, the pair $(L^{(\epsilon)},Q)$ is an admissible pair for the Brownian path $W$ as defined in \autoref{defi:decomp}. \\
-- We now introduce another technical assumption which is a manifestation of the general principle in $v)$ that the vicinity of $x=0^+$ is the only place where the $a$-term plays a significant role: we assume that $\frac{dh_1}{dq}:=\frac{\partial_x h_1(x)}{\partial_x q(x)}$ is bounded when $q$ is bounded away from $0$, or equivalently the function $\frac{a(x)}{2c^2(x)}e^{-2h_0(x)}$ is bounded when $x$ is bounded away from $0$. Letting $c_{\delta}:=\sup \{\frac{dh_1}{dq},q > \delta\}$, we have that $(L^{(\epsilon)},Q)$ is an $(\epsilon c_{\delta},\delta)$  approximate decomposition of $W$. We may adjust $\epsilon$ such that both $\epsilon c_{\delta}$ and $\delta$ are arbitrarily small, and we may conclude that for $\epsilon \to 0^+$ the pair $(L^{(\epsilon)},Q)$ converges uniformly towards the Skorokhod decomposition of $W$ on every compact time interval $[0,T]$. Thus in the limit $\epsilon \to 0^+$, $Q_\tau$ becomes a reflected Brownian motion $|\tilde W_\tau|$ and $L^{(\epsilon)}_\tau$ becomes $L_\tau$, the local time of $\tilde W_\tau$ at the origin.\\
-- Up to now, the discussion has been rigorous. With the particular scale invariant SDEs studied before we could rely on a simple application of the ergodic theorem to prove the crucial relation between the time $t$ of the process $X_t$ and the local time $L_\tau$ of the Brownian motion. This does not work anymore, hence the use of the term ``conjecture''. The approximate local time is related to the physical time via $dL_\tau^{(\epsilon)}=\frac{\lambda^2\epsilon}{2}\, e^{2h_0(X_t)}\,  a(X_t)\, dt$. Since the natural time scale of the problem $\lambda^{-2}$ goes to zero in the scaling limit, we expect that the scaling limit of this integral can be evaluated by first averaging with respect to the invariant measure. That is, we expect that: 
\[ dL_\tau= \lim_{\lambda\to \infty} \frac{\lambda^2\epsilon}{2}\, \langle e^{2h_0(X_t)}\,  a(X_t)\rangle_\mathrm{inv}\, dt .\]
Since the invariant measure is $P_\mathrm{inv}(x)dx= \frac{1}{Z_\epsilon}\, \frac{dx}{c^2(x)}\, e^{-2(h_0(x)+\epsilon h_1(x))}$, we have, using $dh_1(x) =- \frac{ a(x)}{2c^2(x)}dx$:
\[ \langle e^{2h_0(X_t)}\, a(X_t)\rangle_\mathrm{inv} = \frac{1}{Z_\epsilon}\, \int \frac{dx}{c^2(x)} a(x)e^{-2\epsilon h_1(x)}= \frac{2}{Z_\epsilon} \int_0^\infty dh e^{-2\epsilon h} = \frac{1}{\epsilon Z_\epsilon} \]
because by assumption $h_1(x)$ goes to $+\infty$ at small $x$ and to $0$ at large $x$. 
Hence, the relation in the limit is $dL_\tau = \frac{\lambda^2}{2Z_\epsilon}\, dt = \hat J\, dt$. It is quite remarkable that, in the general set-up, this averaging procedure yields an explicit expression of the local time in terms of the scaling variable $\lambda^2/Z_\epsilon$ only.

Similar arguments apply to the reconstruction of the spike point process via passage times:

\begin{conj}
(i) Let $T_{y\to z}$ be the passage time from $y$ to $z>y$ of the SDE (\ref{eq:sde}). In the scaling limit $\lambda\to\infty$, $\epsilon\to 0$, with $\lambda^2/Z_\epsilon=2\hat J$ fixed, its generating function is:
\[ \lim_{\text{scaling}}\mathbb{E}[e^{-\sigma T_{y\to z} } ] = \frac{ 1 + \sigma\, \hat J^{-1}\, q(y)}{1 + \sigma\, \hat J^{-1}\,q(z)}, 
\quad \mathrm{with}\ q(x) = \int_0^x du\, e^{2h_0(u)}.\]
(ii) The spikes of the $X$-trajectories, scaling limits of the solutions of \autoref{eq:sde}, form a point Poisson process on $\mathbb{R}\times\mathbb{R}_+$ with intensity $d\nu = \hat J dt\, \frac{dq(x)}{q(x)^2}$.
\end{conj}

Note that $h_0(x)$ is defined up to an additive constant, so that $q(x)$ is defined up to a multiplicative constant. But so is $\hat J$ and the ratio $q(x)/\hat J$ is invariant (i.e. it is independent of this arbitrary multiplicative constant).

Let us argue in favor of the conjecture. As arguments are close to the ones given in the simpler scale invariant cases, we concentrate on the most delicate analysis, that of the passage time $T_{y\to z}$ for $0 < y < z$.\\
-- We look at the generating function of the passage time $T_{y\to z}$. As before, we have representations:
\[ \mathbb{E}[e^{-\sigma T_{y\to z} } ] = e^{-\int_y^z \phi(x,\sigma)dx} = \frac{\psi(y,\sigma)}{\psi(z,\sigma)},\]
with $\phi(x,\sigma)=\partial_x \log \psi(x,\sigma)$. The function $\phi$ satisfies a non-linear first order differential equation, namely
\[c^2(x)\, \big( \phi'(x) + \phi^2(x)\big) + \big(  \epsilon  a(x) -  b(x)\big)\,\phi(x) = 2\sigma/\lambda^2,\] while the function $\psi$ satisfies a second order linear differential equation $c^2(x)\, \psi''(x) + \big( \epsilon  a(x) -  b(x)\big)\,\psi'(x) = (2\sigma/\lambda^2)\, \psi(x)$. Though we do not indicate it explicitly, $\phi$ and $\psi$ do depend on $\sigma$,  $\epsilon$, $\lambda$.  \\
-- As before, we rewrite the equation for $\phi$ as an integral equation:
for $y, z \in ]0,+\infty [$
\begin{eqnarray} \label{eq:integralform}
\phi(z) e^{-2\epsilon h_1(z)-2h_0(z)} &-& \phi(y) e^{-2\epsilon h_1(y)-2h_0(y)} \nonumber \\ &+ & \int_y^z du \, \phi(u)^2e^{-2\epsilon h_1(u)-2h_0(u)} =  \frac{2\sigma}{\lambda^2} \int_y^z du \,\frac{e^{-2\epsilon h_1(u)-2h_0(u)}}{c(u)^2}.
\end{eqnarray}
One can show from this equation solely that for fixed $\epsilon >0$, $\lim_{y\to 0^+} \phi(y) e^{-2\epsilon h_1(y)-2h_0(y)}=0$ and $\int_0^z du \, \phi(u)^2e^{-2\epsilon h_1(u)-2h_0(u)}$ is finite. We postpone the proof after the main line of argument in an independent lemma. 
This leads to the formula incorporating boundary conditions at $0$: 
\[ 
\phi(z) e^{-2\epsilon h_1(z)-2h_0(z)}+\int_0^z du \, \phi(u)^2e^{-2\epsilon h_1(u)-2h_0(u)}=\frac{2\sigma}{\lambda^2} \int_0^z du \,\frac{e^{-2\epsilon h_1(u)-2h_0(u)}}{c(u)^2}=\frac{2\sigma Z_\epsilon}{\lambda^2}P_\mathrm{inv}(\epsilon,[0,z]).
\] 
We can now investigate the small $\epsilon$ behavior. \\
-- The expectation of the passage time $T_{y\to z}$ is related to the first order in the expansion of $\phi(z)$ in powers of $\sigma$, i.e. to the linearization of the above equation. Quantitatively
\[
\frac{\partial \mathbb{E}[T_{y\to z}]}{\partial z}e^{-2\epsilon h_1(z)-2h_0(z)}=\frac{2}{\lambda^2} \int_0^z du \,\frac{e^{-2\epsilon h_1(u)-2h_0(u)}}{c(u)^2}=\frac{2 Z_\epsilon}{\lambda^2}P_\mathrm{inv}(\epsilon,[0,z]).
\]
Remember our aim is to find a scaling for which $T_{y\to z}$ --in particular its expectation-- remains finite when $\epsilon \to 0^+$. As $\lim_{\epsilon \to 0^+}P_\mathrm{inv}(\epsilon,[0,z])=1$ for each $z>0$, the only way this can occur is by adjusting $Z_\epsilon \lambda^{-2}$ to have a finite limit. This leads to define the scaling limit as the limit $\lambda\to\infty$, $\epsilon\to 0^+$, with $\lambda^2/Z_\epsilon=$ going to a finite limit.\\
-- From now on we fix the quantity $\lambda^2/Z_\epsilon=:2\hat J$ and use this relation to eliminate $\lambda$ from the equations. Thus the scaling limit appears as  $\epsilon\to 0^+$ and $J$ fixed.\\
In particular, we infer from 
 \[ \mathbb{E}[T_{y\to z}] =\frac{1}{\hat{J}}\int_y^z du \, P_\mathrm{inv}(\epsilon,[0,z]) e^{2\epsilon h_1(u)+2h_0(u)}.\]
 that, for $0 < y \leq z$ ,
\[\lim_{\text{scaling}} \mathbb{E}[T_{y\to z}] =\frac{q(z)-q(y)}{\hat J}\]
which will also be a simple consequence of our final result. Then $\lim_{y\to 0^+} \lim_{\text{scaling}} \mathbb{E}[T_{y\to z}]=\hat{J}^{-1} q(z)$. Note that it is unclear whether the limits can be interchanged, i.e. whether this is also $\lim_{\text{scaling}} \mathbb{E}[T_{0\to z}]$. \\
-- The variance of first passage times is related to the next term in the small $\sigma$ expansion, explicitly
$\frac{\partial \mathbb{V}ar[T_{y\to z}]}{\partial z}=\hat{J}^{-2}\int_0^z du \, P_\mathrm{inv}(\epsilon,[0,z])^2 e^{2\epsilon h_1(u)+2h_0(u)}$. 
Note the similarity with the formula for $\mathbb{E}[T_{0\to z}]$ (not its $z$-derivative!) where
the only difference is the power to which $\hat{J}^{-1}P_\mathrm{inv}(\epsilon,[0,z])$ is raised. As $P_\mathrm{inv}(\epsilon,[0,z])$ is small at small $z$ for fixed $\varepsilon >0$, $\frac{\partial \mathbb{V}ar[T_{y\to z}]}{\partial z}$ is well-behaved at small $z$ for fixed $\varepsilon$. We are now in position to make assumption $ix)$ quantitative. We make the technical assumption that
\begin{equation} \label{eq:varbound} \lim_{z \to 0^+} \lim_{\epsilon \to 0^+} \int_0^z du \, P_\mathrm{inv}(\epsilon,[0,z])^2 e^{2\epsilon h_1(u)+2h_0(u)}=0.
\end{equation} 
This condition may be checked in our explicit examples. It means that no region close to $0$ whose size goes to $0$ with $\epsilon$ makes a  noticeable contribution to the integral when $\epsilon \to 0^+$. It is an expression of a regularity condition on the variance, namely 
\[ \lim_{z \to 0^+} \lim_{\text{scaling}} \frac{\partial \mathbb{V}ar[T_{y\to z}]}{\partial z}=0.\]
-- Returning to the global study of $\phi$, We are led to study the small $\epsilon$ limit of the equation 
\begin{equation} \label{eq:phibound}
\phi(z) e^{-2\epsilon h_1(z)-2h_0(z)}+\int_0^z du \, \phi(u)^2e^{-2\epsilon h_1(u)-2h_0(u)}=\frac{\sigma}{\hat J}P_\mathrm{inv}(\epsilon,[0,z]).
\end{equation} 
It involves three terms, and if two terms have a limit when $\epsilon \to 0^+$, namely $\phi(z)$ and $\frac{\sigma }{\hat J}P_\mathrm{inv}(\epsilon,[0,z])$ the third, $\int_0^z du \, \phi(u)^2e^{-2\epsilon h_1(u)-2h_0(u)}$ must have a limit too. The only question is whether the integral of the limit is the limit of the integral. And the only place where a problem may occur is the origin : it could happen that region close to $0$ whose size goes to $0$ with $\epsilon$ makes a contribution to $\int_0^z du \, \phi(u)^2e^{-2\epsilon h_1(u)-2h_0(u)}$ that does not vanish when $\epsilon \to 0^+$. But both terms on the left hand-side of \autoref{eq:phibound} are non-negative, so $\phi(z) e^{-2\epsilon h_1(z)-2h_0(z)}\leq \frac{\sigma }{\hat J}P_\mathrm{inv}(\epsilon,[0,z])$, which leads to 
\[\int_0^z du \, \phi(u)^2e^{-2\epsilon h_1(u)-2h_0(u)} \leq \left(\frac{\sigma }{\hat J}\right)^2 \int_0^z du \, P_\mathrm{inv}(\epsilon,[0,z])^2 e^{2\epsilon h_1(u)+2h_0(u)}. \]
Thus if the condition \autoref{eq:varbound} is fulfilled,  no region close to $0$ whose size goes to $0$ with $\epsilon$ makes a  noticeable contribution to the integral and the limiting function $\phi(z)$ solves the equation 
\[ 
\phi(z) e^{-2h_0(z)}+\int_0^z du \, \phi(u)^2e^{-2h_0(u)}=\frac{\sigma}{\hat J},
\]
whose solution is elementary:
\[ 
\phi(z)=\frac{\sigma}{\hat J} \frac{q'(z)}{1+\frac{\sigma}{\hat J}q(z)} \quad q(z)=\int_0^z  du \, e^{2h_0(u)}
\]
leading to the announced formula for $\lim_{\text{scaling}}\mathbb{E}[e^{-\sigma T_{y\to z} } ]$.\\
-- As  a final consistency check, let us observe that the function $\phi $ in the scaling limit, which is of the form of an outer solution,  is linear in $\sigma$ at small $z$ and reads $\phi(z) \simeq \frac{\sigma}{\hat J} e^{2h_0(z)}$, to be compared with the complete (finite $\epsilon$) solution to first order in $\sigma$, which reads 
$\phi(z) \simeq \frac{\sigma}{\hat J} e^{2h_0(z)} e^{2\epsilon h_1(z)} P_\mathrm{inv}(\epsilon,[0,z])$. Thus there is matching if there is some ($\epsilon$-dependent) region of the variable $z$ for whom $e^{2\epsilon h_1(z)} P_\mathrm{inv}(\epsilon,[0,z])\simeq 1$. This relation is known to hold for fixed $z$ and small $\epsilon$, and this is enough to guarantee the existence of a matching region, though its precise form depends on the details of the model. 

Once the formula $\lim_{\text{scaling}}\mathbb{E}[e^{-\sigma T_{y\to z} } ] = \frac{ 1 + \sigma\, \hat J^{-1}\, q(y)}{1 + \sigma\, \hat J^{-1}\,q(z)}$ is established, the construction of the process of maxima and the rest of the discussion can be copied word for word from \autoref{ssec:ppmsl}.

We finish the discussion with the lemma needed to take the limit $y\to 0^+$ in \autoref{eq:integralform}:
\begin{lem}
	Suppose $f(x)$ is a function defined for $x$ large enough, non-negative and such that $\lim_{x \to +\infty} \int^x f^2(y)dy-f(x)$ exists and is finite. Then $\lim_{x \to +\infty} f(x)=0$ and $\int^{+\infty} f^2(y)dy$ is finite.
\end{lem} 

{\bf Proof:} As $\int^x f^2(y)dy$ is an increasing function of $x$, it has a non-negative limit, finite or infinite, when $x\to +\infty$. Hence, by hypothesis, the same has to be true of $f(x)$. If the limit of $f(x)$ is finite and non-zero, then obviously $\int^x f^2(y)dy$ diverges at large $x$, contradicting the existence of a limit for $\int^x f^2(y)dy-f(x)$. Thus is remains only to exclude the possibility that $f(x)$ goes to $+\infty$ when $x\to +\infty$, and we argue by contradiction. By hypothesis, there is a constant $C$ such that $\int^x f^2(y)dy-f(x) \leq C$ for large enough $x$, and if $\lim_{x \to +\infty} f(x)=+\infty$ then $lim_{x \to +\infty}\int^x f^2(y)dy =+\infty$ so that for $x$ large enough $0 < \int^x f^2(y)dy -C $. Thus setting $g(x):=\int^x f^2(y)dy -C$ there is an $x_0$ such that $0 \leq g(x) \leq f(x)$ for $x \geq x_0$. But $g$ is differentiable and $g'(x)=f^2(x)$ which is $\geq g(x)^2$ for $x \geq x_0$. Thus $\frac{g'(x)}{g(x)^2}\geq 1$ for $x \geq  x_0$. Taking the integral leads to $\frac{1}{g(x_0)}-\frac{1}{g(x)} \geq x-x_0$ for $x \geq  x_0$. But the left-hand side has a finite limit at large $x$ because $g(x)$ is large at large $x$ so this is again a contradiction. $\square$

\begin{cor}
	Let $e(z), f(z)$ are non-negative functions defined in a neighborhood of $O^+$ and such that $\int_z e(u)du$ diverges when $z \to 0^+$ and $\lim_{z \to 0^+} \int_z f^2(u)e(u)du-f(z)$ exists and is finite. Then $\lim_{z \to 0^+} f(z)=0$ and $\int_{0^+} f^2(u)e(u)du$ is finite. 
\end{cor}

{\bf Proof:} Setting $x(z)=\int_z e(u)du$ and changing variables from $z$ to $x$, we are reduced to an application of the lemma. $\square$

We apply the corollary to \autoref{eq:integralform}. We set $e(z):=e^{2\epsilon h_1(z)+2h_0(z)}$ and  $f(z):=\phi(z) e^{-2\epsilon h_1(z)-2h_0(z)}$ so that the equation becomes 
\[ f(z)-f(y)+\int_{y}^{z} f^2(u)e(u)du =\frac{2\sigma}{\lambda^2} \int_y^z du \,\frac{e^{-2\epsilon h_1(u)-2h_0(u)}}{c(u)^2}.\]
By assumption $vi)$, $\int_y e(u)du$ diverges when $y \to 0^+$, while by assumption $vii)$ the integral on the right-hand side has a limit when $y \to 0^+$. Hence the corollary implies that $\lim_{y\to 0^+} \phi(y) e^{-2\epsilon h_1(y)-2h_0(y)}=0$ and $\int_0^z du \, \phi(u)^2e^{-2\epsilon h_1(u)-2h_0(u)}$ is finite, as needed in the main line of the argument. 

\subsection{An example}

Let us consider the case $b(x) = bx$ and $c(x)=x^2$ which  for $b=1$ corresponds to the homodyne detection of Rabi oscillation (see \autoref{sec:MQ}). The SDE is then:
\beq \label{sde:ex}
 dX_t = \frac{\lambda^2}{2}( \epsilon -  bX_t)\, dt + \lambda\, X_t^2\, dB_t.
 \eeq
It is not scale invariant. Its invariant measure is 
\[ P_\mathrm{inv}(x)dx = \frac{1}{Z_\epsilon} \frac{dx}{x^4} e^{+b/2x^2-\epsilon/3x^3}.\]
The normalization factor is $Z_\epsilon:= \int_0^\infty \frac{du}{u^4}\, e^{\frac{b}{2}u^{-2}-\frac{\epsilon}{3}u^{-3}}$ which, via a change variable, can be written as
\[ Z_\epsilon=\frac{1}{\epsilon^3} \int_0^\infty ds\, s^2\, e^{(\frac{bs^2}{2}-\frac{s^3}{3})/\epsilon^{2}}.\]
In the limit $\epsilon\to 0$ this integral is dominated by the saddle point at $s=b$. Hence
\[ Z_\epsilon \simeq_{\epsilon\to 0}  \sqrt{2\pi b^3}\, \epsilon^{-2}\, e^{b^3/6\epsilon^2}  .\]
It is not a simple power law in $\epsilon$ because the SDE (\ref{sde:ex}) is not scale invariant. The relation defining the scaling limit, ${\lambda^2}/{2Z_\epsilon}=:\hat J$ fixed, reads
\[ {(\lambda^2\epsilon^2)\, e^{-b^3/6\epsilon^2} }= \hat J\, \sqrt{8\pi b^3}\quad \mathrm{fixed} ,\]
as $\epsilon\to 0$, $\lambda\to \infty$. It is non-perturbative in $\epsilon$ and bears similarities with spontaneous mass generation in asymptotically free field theory. 

The change of variable $X\to Q$ needed to extract the reflected Brownian motion cannot be made explicit (because of the absence of scale invariance of \autoref{sde:ex}) but it reads:
\[ Q_t:= q(X_t) = \int_0^{X_t} du\, e^{-b/2u^2}.\]
$Q_t$ is a local martingale in absence of the $\epsilon$ term, otherwise $dQ_t = \frac{\lambda^2\epsilon}{2} e^{-b/2X_t^2}\, dt + \lambda\, X_t^2\, e^{-b/2X_t^2}\, dB_t$. The effective time $\tau$ and the corresponding Brownian motion $W_\tau$ are defined by
\[ dW_\tau = \lambda\, X_t^2\, e^{-b/2X_t^2}\, dB_t,\quad d\tau = \lambda^2\, X_t^4\, e^{-b/X_t^2}\, dt.\]
We then apply the Skorokhod's theorem and get that, in the scaling limit, $Q_\tau$ is a reflected Brownian motion, 
\[ Q_\tau=|\tilde W_\tau|,\]
and $dQ_\tau = dL_\tau + dW_\tau$ with $L_\tau$ the $\tilde W_\tau$ local time at the origin. 

Because of the lack of scale invariance we cannot use the ergodic theorem to restore the relation between the effective local time and the original time. We instead make use of the averaging trick (for which we do not have a formal proof):
\[ dL_\tau = \frac{\lambda^2\epsilon}{2} \langle e^{-b/2X_t^2} \rangle_\mathrm{inv} \, dt .\]
Since the invariant measure is $\frac{1}{Z_\epsilon} \frac{dx}{x^4} e^{+b/2x^2-\epsilon/3x^3}$, we have 
$\langle e^{-b/2X_t^2} \rangle_\mathrm{inv} =  \frac{1}{Z_\epsilon} \int_0^\infty \frac{dx}{x^4} e^{-\epsilon/3x^3}= \frac{1}{\epsilon Z_\epsilon}$. 
Finally
\[ dL_\tau = \frac{\lambda^2}{2Z_\epsilon}\, dt = \hat J\, dt .\]
We thus recover, by construction, the expected time reconstruction relation.

\section{Small noise limit: Comparison}
\label{sec:weak}

In this short Section, we would like to compare the large noise spiky behavior studied in this article with the behavior of solutions in the well-known small noise limit, see e.g. the book \cite{FW}. First note that the existence of spikes in the strong noise limit is linked to the heavy tail of the invariant measure which is linked to the presence of multiplicative noise with fixed points, i.e. non-triviality of $c(x)$.

To simplify matters we look at a SDE for a single variable of the form\footnote{We could also have considered small noise SDEs with multiplicative noise but this would not change the discussion much.}:
\beq \label{sde:small}
 dX_t = - U'(X_t) dt + \nu dB_t,
 \eeq
where $\nu$ controls the strength of the noise. The small noise limit is for $\nu\ll 1$. 
To fix the setup we take $U(x)$ to be a double well potential, infinite at $x_{\pm\infty}$, with two minima at $x_0$ and $x_1$ and a local maximum at $x_*$. The convention is that $x_{-\infty}<x_0<x_*<x_1<x_{+\infty}$.

The two minima are metastable attractive points but the presence of noise induces Kramer's like transitions from one to the other. The aim of this Section is to compare these jumps and the associated structure of aborted jumps in the small noise limit $\nu\to 0$ with the large noise spiky structure discussed in the previous Sections.

Let $T_{y\to z}$ be the time to go from $y$ to $z>y$ for \autoref{sde:small}. As is well known, down-hill processes, i.e. those for $U(z)<U(y)$, are classical in the small noise limit. They are dominated by the drift, so that $dX_t\simeq -U'(X_t)dt$, or $dt \simeq -{dX_t}/{U'(X_t)}$, and $T_{y\to z}$ is asymptotically deterministic with $T_{y\to z}\simeq \int_y^z \frac{dx}{U'(x)}$.
Up-hill processes, i.e. for $U(z)>U(y)$, are highly non-classical. They are induced by the noise and are non-perturbative in $\nu$. They are typically exponentially large: $T_{y\to z}\simeq e^{+2(\Delta_{y\to z} U)/\nu^2}$, up to multiplicative sub-exponential prefactor, with $\Delta_{y\to z} U:= U(z)-U(y)$.

More precisely, let $\bar T_{y\to z}$ be the mean first passage time. By the strong Markov property we have $\bar T_{y\to z}=\int_y^zdx\, \phi_1(x)$. One can prove (e.g by introducing the generating function $ \mathbb{E}[e^{-\sigma T_{y\to z} } ]$ and analyzing the differential equation it satisfies, see \cite{FW}) that near one of the minima, say $x_0$, we have asymptotically as $\nu\to0^+$: 
\[ \phi_1(x) = 
\begin{cases}  
-{1}/{U'(x)} &,\ x_{-\infty}<x<x_\nu<x_0\\ \\
\nu^{-1}\sqrt{\frac{2\pi}{U''(x_0)}}\, e^{ +2(\Delta_{x_0\to x}U)/\nu^2} &,\ x_0<\tilde x_\nu< x <x_*
\end{cases}
\]
These two formula are valid except in a small interval $[x_\nu,\tilde x_\nu]$  around the minimum $x_0$. 
They also have to be modified when $x$ approaches the saddle maximum $x_*$. There the mean first passage time is still exponentially large, say $\bar T_{x_0\to x_*}\simeq e^{+2(\Delta_{x_0\to x_*} U)/\nu^2}$, but the multiplicative prefactor is different. 

\begin{figure}
	\begin{center} \includegraphics[width=\textwidth]{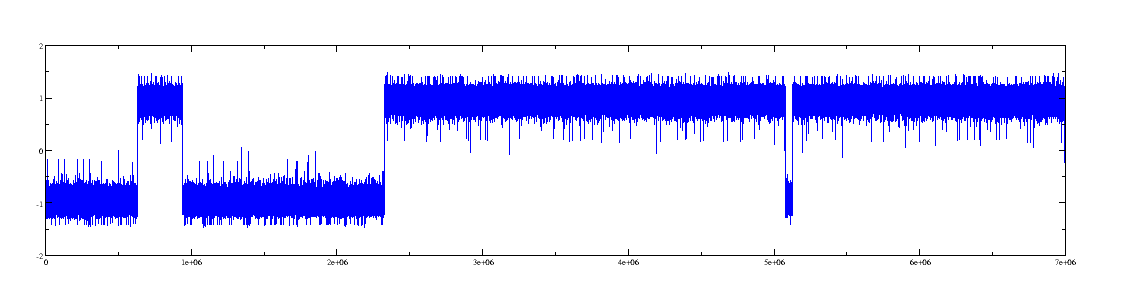} \end{center}
	\caption{\emph{A weak noise simulation}}
	\label{fig:smallnoise}
\end{figure}

Since these first passage times diverge, we have to fix a point $x_\bullet$ up-hill from $x_0$ $(x_0<x_\bullet<x_*)$ and rescale time, in order to have a meaningful limit $\nu\to0$.  That is, we scale down time, so that the mean time to go from $x_0$ to $x_\bullet$ is finite in the small $\epsilon$ limit, and define a new time variable $t'$ via
\[ t \to t':= t/\bar T_{x_0\to x_\bullet} \]
so that $\bar T'_{x_0\to x_\bullet}=1$. We then have a dychotomic behavior in the limit $\nu\to0$:
\[ \bar T'_{x_0\to x}= \begin{cases} 
0 &,\ \mathrm{if}\ x_0<x<x_\bullet\\
\infty &,\ \mathrm{if}\ x_\bullet<x<x_* 
\end{cases} \]
As a consequence the picture of the graph trajectories is quite different from the strong noise limit. In the small noise limit:\\
-- The depth of the different wells has to be fine-tuned if all of them have to lead to commensurable time scales so that the jumps between them can be observed on a single time scale. This is clearly visible even in small but finite noise simulations.\\
-- When looked at in a time scale making jump times finite, the graph of the trajectory looks like a succession of filled rectangles joined by a corner. Rectangular shapes also show up in small but finite noise simulations, but they are joined by finite vertical segments and perturbed by rough bumps, see \autoref{fig:smallnoise}. 
The reasons are twofold. First, what can be implemented as small noise in numerical simulations is not so small. As an illustration, for the double well with $U(x):= (x^2-1)^2/4$ (so the depth of the wells is $1/4$) and $\nu=1/5$, the inter-jump time time is of order $10^6$ (see \autoref{fig:smallnoise}). When $\nu=1/6$, the inter-jump time time is already of order $10^8$, to be compared to a down-hill time scale of order $10^0$. Discretized with a time step of order $10^{-3}$, the simulation takes roughly $10^{12}$ time steps, i.e. about one day on a laptop. But taking $\nu=1/7$ would take weeks. Second, keeping all the computed values, at $\nu=1/6$ say, would take about $10^2$TB of data, so only a small fraction can be kept for the picture. As a consequence, accessible simulations are still quite far from the zero noise limit and they are not faithful.

There is however one salient common point with the large noise limit studied in this article: the time separating two successive passages at $x_0$ and $x_1$ is exponentially distributed in the small noise limit, just like the time between two quantum jumps in the large noise limit. But in the large noise limit, the same time scale governs also first passage times at intermediate points, whereas in small noise equations, any $x \in [x_0<x_* [$ is visited an exponentially large number of times times before there is a jump from $x_0$ to $x_1$, and in a time scale where the first passage time from $x_0$ to $x_1$ is finite, the first passage time from $x_0$ to any $x \in [x_0<x_* [$ is exponentially small. At small but finite noise, the occupation times within a well in-between two jumps are to an excellent approximation proportional to the stationary measure, accounting for the bumps visible on \autoref{fig:smallnoise}. 

\section{Discussion}

In this paper, we have argued, and proven in a number of important special cases, that generically solutions of stochastic differential equations with a fixed point subject to an  additional infinitesimal repulsive perturbation exhibit a universal behavior when time is rescaled appropriately: by tuning the time scale with the infinitesimal repulsive perturbation to ensure a finite first passage time, the trajectories converge in a precise sense to spiky trajectories that can be reconstructed from an auxiliary time-homogeneous Poisson process. The rescaling of the time scale is a large noise limit, but combined with a rescaling of the repulsive perturbation to compensate for the quantum Zeno effect.   

Our results are based on two main tools. The first is a time change and an application of Skorokhod's lemma. To reach our goal, we have proven an effective approximate version of this lemma which is new as far as we are aware of and which is of independent interest. The second is an analysis of first passage times, which shows a nice interplay between scale functions and invariant measures.  

We have also stressed the profound difference with standard weak noise limits (Kramer's theory in the physics language) 

Our results raise a number of questions, and we list a few:\\
-- What happens when more than one fixed point exists? We have convincing evidence (and sometimes proofs, see e.g. \cite{BBTskospikes}) that the main ideas remain valid, the role of the Poisson process being played by a Cox process. \\
-- What happens when several components are present? Numerical simulations and hand-waving arguments suggest that a spiky behavior with universal features is still the rule, but the precise description of the associated Poisson (or Cox) processes is at the moment completely terra incognita. We hope to return to these questions in the future. \\
-- Are there other areas where these ideas can be applied? There are a number of examples of natural phenomena exhibiting spiky behavior. One can for instance think of some geophysical data, but one very striking case if intermittency in turbulence (see e.g. \cite{Frisch}). A precise connexion would be an extremely exciting and important discovery. But at the moment it is unclear whether the resemblance is deeper than just visual. In the case of quantum mechanics, the time change that simplifies the dynamics is dictated by the need to use not the number of measurements but the effect of measurements on the system to describe the evolution, leading to the notion of effective time. And it turns out that the right way to quantify this effect is by identifying this effective time with the quadratic variation. But this does not have to hold for more general intermittent behaviors, for which one has to decide between different ways of quantifying the notion of effective time before one can make comparisons with ``simpler processes''. This also raises questions about the multi-fractal spectrum of the spiky trajectories studied in this article. We hope to return to these questions as well in the future. \\

\bigskip

\noindent {\bf Acknowledgements}: 
This work was in part supported by the ANR project ``StoQ'', contract number ANR-14-CE25-0003. MB and DB thank Martin Kolb and Antoine Tilloy for discussions. MB thanks Alain Comtet, Patrick Polo and Yves Tourigny for discussions. 
\bigskip

\appendix

\section{An illustrative computation}

We describe in this appendix how some of the results of  \autoref{ssec:ppmsl} can be re-derived assuming the spiky process is a Poisson point process. In particular the limiting distribution of the passage time determine the intensity of this point  process.
The crucial point is that the generating function for the passage times in \autoref{dist:T} is the ratio ${\psi(y,\sigma)}/{\psi(z,\sigma)}$ with $\psi(x,\sigma)$ a  polynomial of degree one in $\sigma$.

Consider a Poisson point process in $\mathbb{R}\times\mathbb{R}^+$ with intensity $d\nu=dt\, d\hat \nu(x)$. To any sample of this process we are going to associate a spiky trajectory $t\to X_t$ (denoting by $X_t$ these trajectories is a slight abuse of notation which is justified by the fact that these trajectories have the same distribution as the solutions $X_t$ of the SDE (\ref{sde:b0bis}) in the scaling limit). The construction is as follows: Pick a sample of the point process and consider all its points above a small thresholds $\eta\to 0^+$. There is a countable numbers of such points: let us enumerate them as $(t_j,x_j)$, $x_j>\eta$, in increasing time order $0<t_1<t_2<\cdots$. The associate trajectories $X_t$ is then defined to be zero on all open interval $]t_{j-1},t_j[$ (i.e. $X_t=0$ for $t\in]t_{j-1},t_j[$) and to be equal to the point of the process in between (i.e. $X_{t_j}=x_j$ for all $j$).

That is: the trajectories are made of series of spikes emerging from the origin at random times and whose tips are the points of the Poisson process. The distribution of these trajectories is induced by that of the point process.
Given such trajectories we can then defined the passage time $T_{y\to z}$ as the time to go from $y$ to $z>y$.

\begin{prop}
\it The distribution for the passage time $T_{y\to z}$ for the spiky trajectories associated to a Poisson point process of intensity $d\nu=dt\, d\hat \nu(x)$ is:
\beq \label{ppp:T}
 \mathbb{P}\big[T_{y\to z}\in [t,t+dt]\big] = \frac{\hat \nu([z,\infty))}{\hat \nu([y,\infty))}\Big( \mathbf{1}_{t=0} + \hat \nu([y,z))e^{-\hat \nu([z,\infty))t}dt\Big)
 \eeq
\end{prop}

{\bf Proof:}
Let $\mathcal{N}_{[t_0,t]\times I}$, with $I$ a Borel set in $\mathbb{R}^+$, be the number of points of the process in $[t_0,t]\times I$. 
There are two contributions to the distribution of the transition times from $y$ to $z$:\\
(i) either the spike going above $y$ at initial time $t=0$ is also going above $z$;\\
(ii) or the spike going above $y$ at initial time $t=0$ stops before reaching $z$ and the first later spike going above $z$ is at time $t$, up to $dt$. \\
The probability of the first event is
\[ \lim_{\delta\to 0} \mathbb{P}\big[ {\cal N}_{[0,\delta]\times [z,\infty)}=1\big\vert\, {\cal N}_{[0,\delta]\times [y,\infty)}=1\big]
=  \frac{\hat \nu([z,\infty))}{\hat \nu([y,\infty))} \]
The probability of the second event is
\[
\begin{split}
& \lim_{\delta\to 0} \mathbb{P}\big[ {\cal N}_{[0,\delta]\times[y,z)}=1,\, {\cal N}_{[0,\delta]\times[z,\infty)}=0,\, {\cal N}_{[t,t+dt]\times[z,\infty)}=1\big\vert\, {\cal N}_{[0,\delta]\times[y,\infty)}=1\big]\\
& \hskip 1.5 truecm =\frac{\hat \nu([y,z))\, \hat \nu([z,\infty))}{\hat \nu([y,\infty))}\, e^{-\hat \nu([z,\infty))t}\, dt
\end{split}
\]
Summing up these two contributions we get the probability that the transition time $T_{y\to z}$ be $t$ up to $dt$ is given by \autoref{ppp:T}.
Remark that this distribution is correctly normalized, thanks to $\hat \nu([y,\infty))=\hat \nu([y,z))+\hat \nu([z,\infty))$.
$\square$

The corresponding generating function $\mathbb{E}[e^{-\sigma T_{y\to z} } ]$ is then computable by integration. We get:
\[ \mathbb{E}[e^{-\sigma T_{y\to z} } ]= \frac{\hat \nu([z,\infty))}{\hat \nu([y,\infty))}\frac{\hat \nu([y,\infty])+\sigma}{\hat \nu([z,\infty))+\sigma} 
=  \frac{1+ \sigma\, \hat \nu([y,\infty))^{-1}}{1+ \sigma\, \hat \nu([z,\infty))^{-1}} , \]
or alternatively, $\psi(x,\sigma)= 1+ \sigma\, \hat \nu([x,\infty))^{-1}$ up to a multiplicative constant. Comparing with \autoref{dist:T} we read that the tips of the $X$-spikes form a point Poisson process with intensity
\[  d\nu = \hat Jdt\, \frac{dq(x)}{q(x)^2} =  \frac{(b+1)}{2\Gamma(b+1)}\, Jdt\,\frac{ dx}{x^{b+2}},\]
as claimed.


\begin{thebibliography}{xxx} 
	
\bibitem{BBTskospikes} M. Bauer, D. Bernard and A. Tilloy,
\textit{Zooming in on Quantum Trajectories},
J. Phys. A: Math. Theor. \textbf{49} (2016), 10LT01, \texttt{ arXiv:1512.02861}.

\bibitem{FW} M.-I. Freidlin, A.-D. Wentzell, \textit{Random Perturbations of Dynamical Systems}, $3^{rd}$ edition, Springer Verlag (2012).

\bibitem{Frisch} U. Frisch, \textit{Turbulence, the legacy of A.N. Kolmogorov}, Cambridge University Press (1995).

\bibitem{ito-mckean}
H. P.~McKean K.~It\^o, \textit{Diffusion Processes and their Sample Paths}, Classics in Mathematics. Springer Verlag, 1991. 

\bibitem{skorokhod} A. V. Skorohod,
\textit{Stochastic equations for diffusion processes in a bounded region 1, 2},
Theor. Veroyatnost. i Primenen. \textbf{6} (1961), 264-274; \textbf{7} (1962), 3-23.

\bibitem{yenyor}
J-Y Yen, M. Yor, \textit{Local Times and Excursion Theory for Brownian Motion},
Number \textbf{2008} in {Lecture Notes in Mathematics}. Springer, 2013.






\end{thebibliography}
\end{document}